\documentclass[
prd,
twocolumn,
aps,
superscriptaddress,
showpacs,preprintnumbers,
amsmath,
amssymb]{revtex4-2}
\usepackage{graphicx}
\usepackage{xspace}
\usepackage{epsfig}
\usepackage{amsfonts}
\usepackage{multirow}
\usepackage{amsmath}
\usepackage{dcolumn}  
\usepackage{wasysym}
\usepackage{multirow}
\usepackage{subfigure}
\usepackage{longtable} 

\usepackage[colorlinks,linkcolor=red,citecolor=blue, bookmarks=false]{hyperref}

\usepackage{orcidlink}
\newcommand{\mb}{\ensuremath{M_{\mathrm{bc}}}\xspace}
\newcommand{\de}{\ensuremath{\Delta E}\xspace}

\newcommand{\acp}{\ensuremath{\mathcal{A}_{\mathit CP}}\xspace}
\newcommand{\br}{\ensuremath{\mathcal{B}}\xspace}

\newcommand{\ks}{\ensuremath{K^0_S}\xspace}
\newcommand{\bb}{\ensuremath{B \bar{B}}\xspace}
\newcommand{\mkk}{\ensuremath{M_{K^+K^-}}\xspace}
\newcommand{\kk}{\ensuremath{{K^+K^-}}\xspace}
\newcommand{\kpi}{\ensuremath{{K^- \pi^+}}\xspace}
\newcommand{\ebeam}{E_\mathrm{beam}}

\newcommand{\Nsig}{N_{\rm sig}}
\newcommand{\kkpi}{\ensuremath{{B^+ \to K^+K^- \pi^+}}\xspace}
\newcommand{\mkpi}{\ensuremath{M_{K^+\pi^-}}\xspace}
\newcommand{\control}{\ensuremath{{B^+ \to \bar{D}^0(K^{+}K^{-}) \pi^+}}\xspace}

\renewcommand{\arraystretch}{1.1}
\def\calL{{\mathcal L}}

\def\BR{{\mathcal B}}
\def\Mbc{M_{\rm bc}}
\def\nn{{\mathcal C_{\mathit NN}}}
\def\cpid{\mathcal{C}_{\mathrm{PID}}}
\def\rkpi{\mathcal{R}_{K/\pi}}

\begin{document}
\preprint{\vbox{
							   \hbox{Belle Preprint 2022-07}
							   \hbox{KEK Preprint 2022-5}
}}

\title{
{\Large \bf \boldmath Angular analysis of the low \kk \textcolor{black}{invariant} mass enhancement in \kkpi decays}
}
\noaffiliation
  \author{C.-L.~Hsu\,\orcidlink{0000-0002-1641-430X}} 
  \author{M.~E.~Sevior\,\orcidlink{0000-0002-4824-101X}} 
  \author{I.~Adachi\,\orcidlink{0000-0003-2287-0173}} 
  \author{H.~Aihara\,\orcidlink{0000-0002-1907-5964}} 
  \author{S.~Al~Said\,\orcidlink{0000-0002-4895-3869}} 
  \author{D.~M.~Asner\,\orcidlink{0000-0002-1586-5790}} 
  \author{H.~Atmacan\,\orcidlink{0000-0003-2435-501X}} 
  \author{V.~Aulchenko\,\orcidlink{0000-0002-5394-4406}} 
  \author{T.~Aushev\,\orcidlink{0000-0002-6347-7055}} 
  \author{R.~Ayad\,\orcidlink{0000-0003-3466-9290}} 
% ...
\collaboration{The Belle Collaboration}

\begin{abstract}

We study the decay $B^{+} \to K^+ K^- \pi^+$ and investigate the angular distribution of $K^{+}K^{-}$ pairs with invariant mass below $1.1$ GeV/$c^2$. This region exhibits both a strong enhancement in signal and very large direct $CP$ violation. We construct a coherent sum model for the angular distribution of the $S$- and $P$-wave, and report the ratio of their amplitudes, the relative phase and the forward-backward asymmetry.
We also report absolute differential branching fractions and direct $CP$ asymmetry for the decay in bins of $\mkk$ and the differential branching fractions in bins of $\mkpi$.
The results are based on a data sample that contains $772\times10^6$ $\bb$ pairs collected at the $\Upsilon(4S)$ resonance with the Belle detector at the KEKB asymmetric-energy $e^+ e^-$ collider. \textcolor{black}{The result 
favours the presence of $S$- and $D$-waves in low $\mkk$ region to the detriment of a $P$-wave.}

\end{abstract}
\maketitle

\section{\label{sec:intro}Introduction}
\begin{figure}[tbp]
\includegraphics[width=0.48\columnwidth]{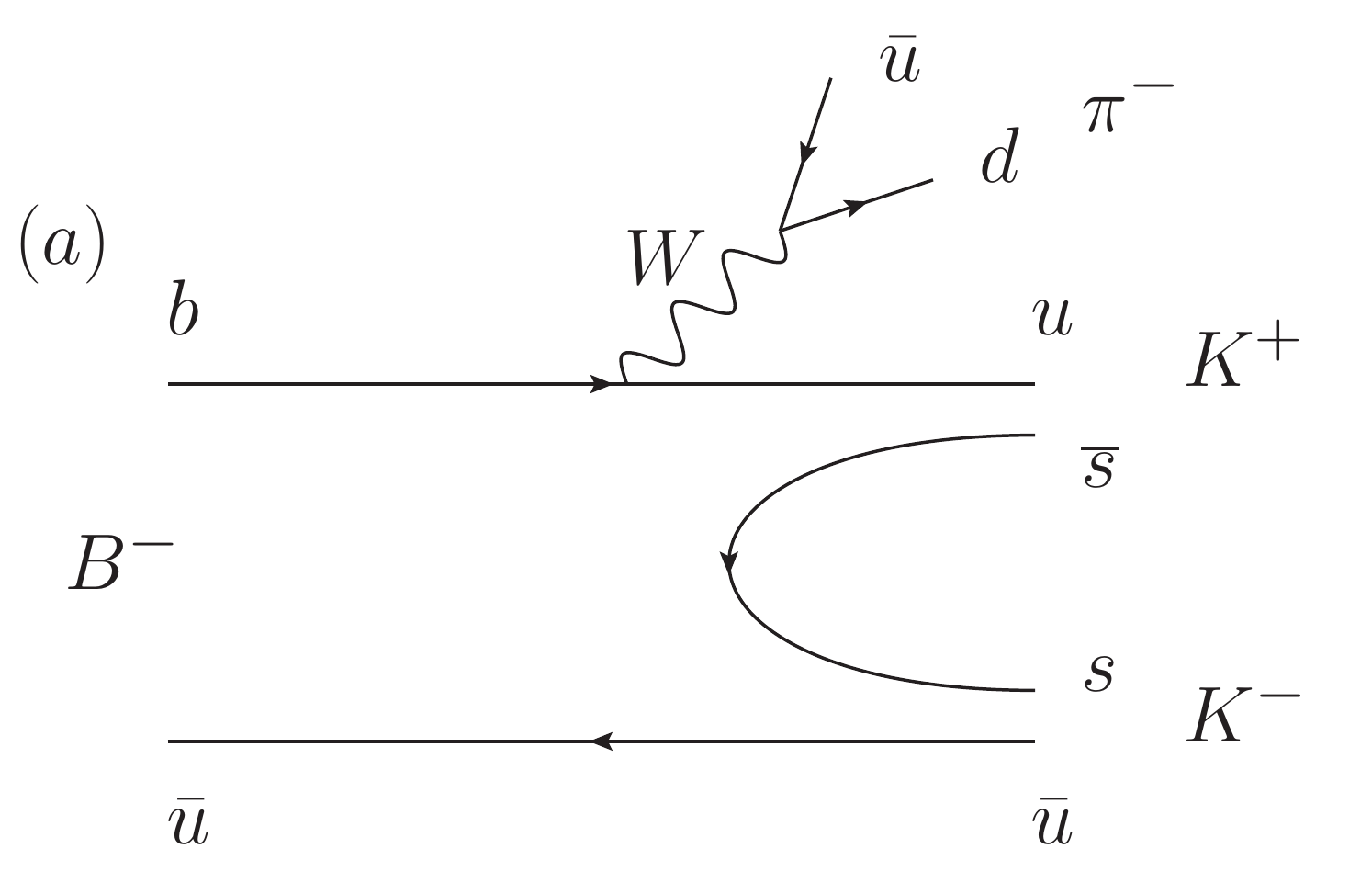}
\includegraphics[width=0.48\columnwidth]{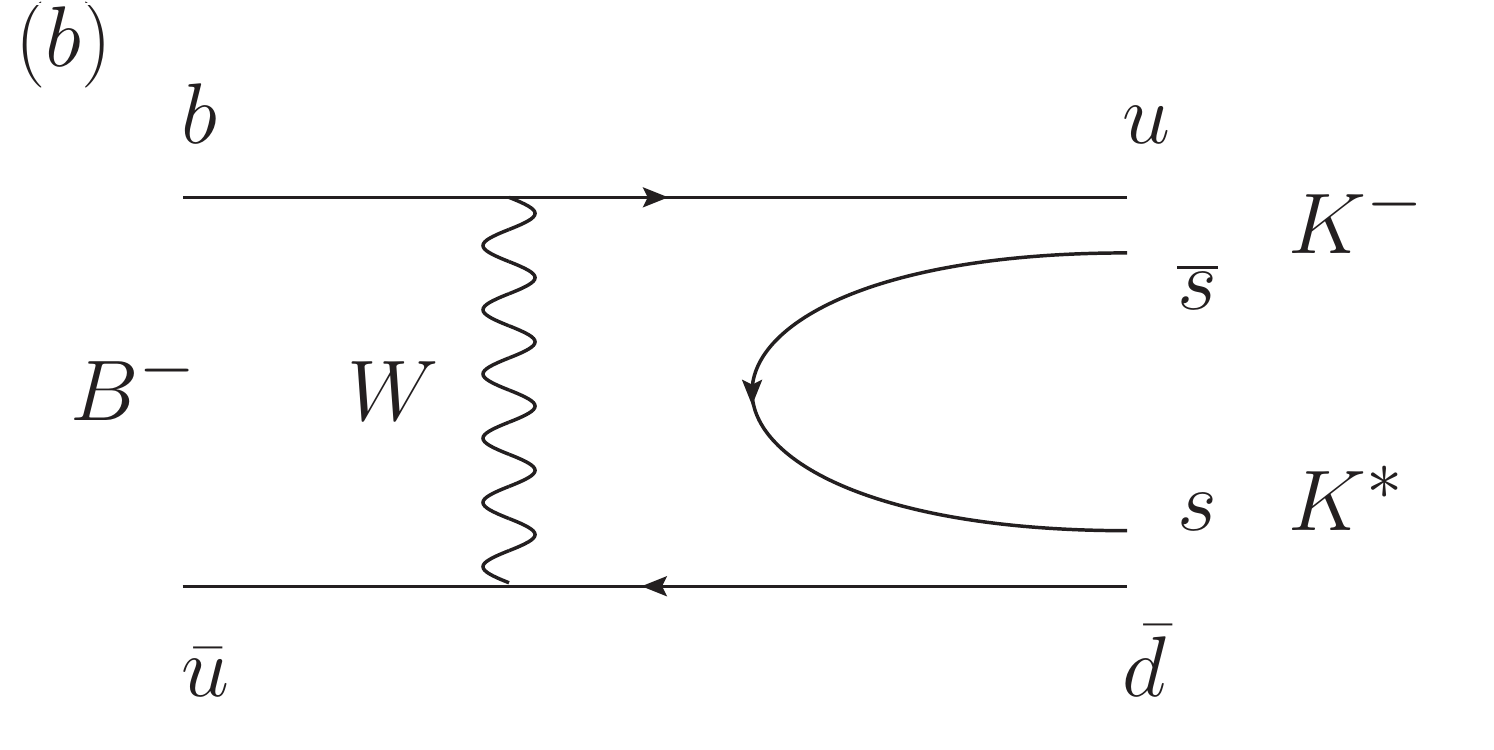}
\includegraphics[width=0.48\columnwidth]{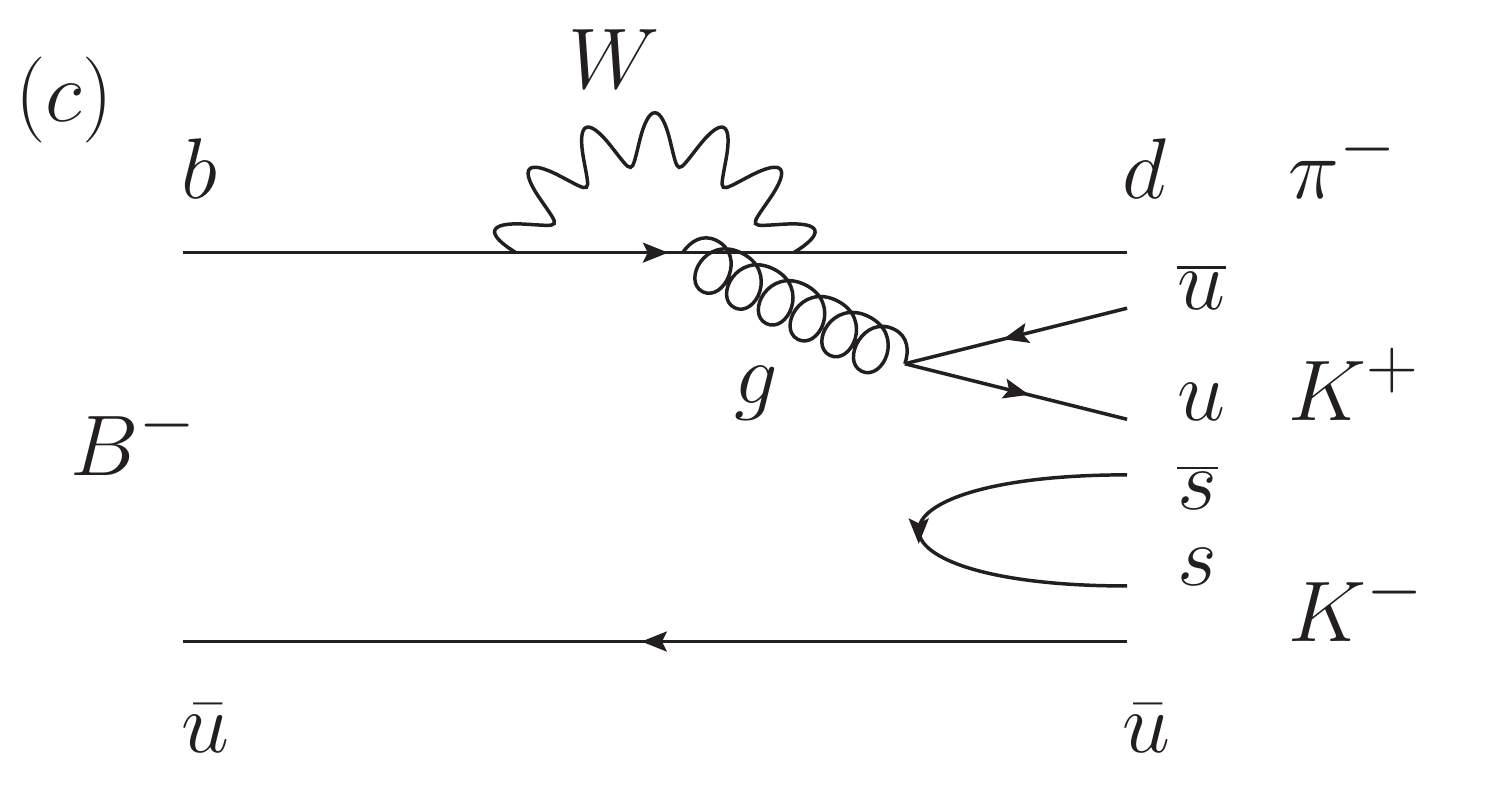}
\includegraphics[width=0.48\columnwidth]{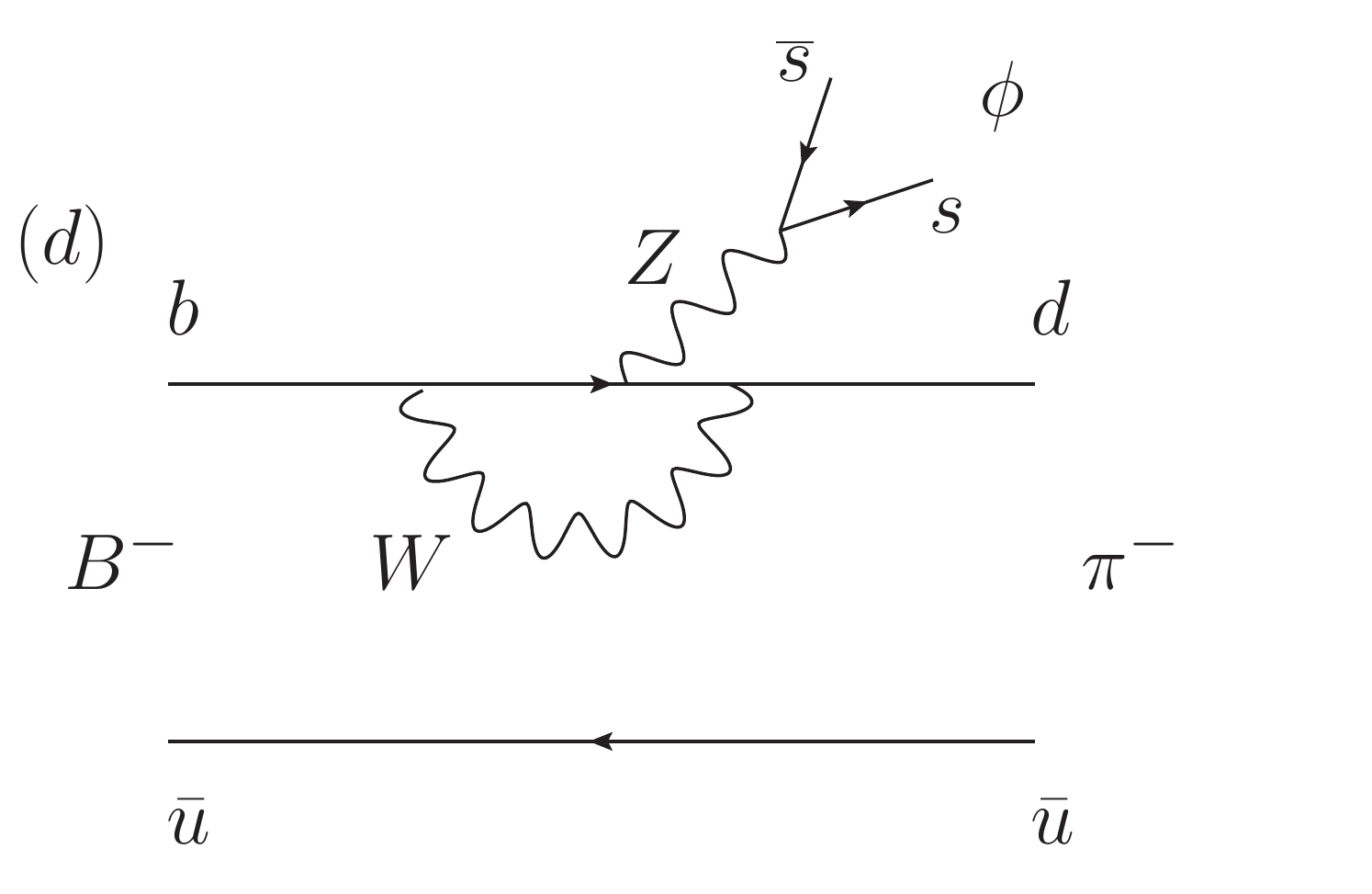} 
\caption{$\kkpi$ Feynman diagrams (all CKM suppressed). (a) Tree diagram, (b) $W$-exchange diagram leading to $KK^{*}$ states, (c) strong-penguin diagram, and (d) electroweak penguin leading to the $\phi\pi$ state.}
\label{fig:fey_dia}
\end{figure}

Charmless decays of $B$ mesons  are suppressed in the Standard Model (SM) and thus provide an opportunity to search for physics beyond the SM through branching fraction enhancements.
Large $CP$ asymmetries can occur in these decays due to interference of SM tree and loop diagrams with similar amplitudes; there is also the possibility of beyond-SM particles contributing in the loop diagrams.
Figure~\ref{fig:fey_dia} shows 
some SM Feynman diagrams that contribute to the $\kkpi$ decay~\cite{conjugate}.
The dominant process is the Cabibbo-suppressed $b \to u$ tree transition in Fig~\ref{fig:fey_dia}(a); 
the $b \to d$ penguin diagram in Fig.~\ref{fig:fey_dia}(d) leading to $B^+ \to \phi \pi^{+}$ with $\phi \to \kk$ is heavily suppressed, and the current experimental limit for this mode is $\BR(B^+\to\phi\pi^+)<1.5\times 10^{-7}$~\cite{lhcb_phipi}.

An unidentified structure has been observed by BaBar~\cite{BaBar_kkpi} and LHCb~\cite{lhcb_old,lhcb_new,lhcb_dalitz} in the 
$\kk$ low-invariant-mass spectrum of the $\kkpi$ decay. The LHCb studies also found a nonzero \textcolor{black}{integrated} $CP$ asymmetry of $-0.123\pm0.022$
and a large local $CP$ asymmetry in the same mass region.
References~\cite{Bhattacharya2013,Bediaga2014} suggest that final-state interactions may enhance $CP$ violation and the recent LHCb study suggests that the large $CP$ asymmetry in the low-$\kk$ invariant-mass region originates from $\pi\pi \leftrightarrow KK$ $S$-wave rescattering~\cite{lhcb_dalitz}.
Building on our earlier shorter publication~\cite{Hsu}, this paper studies the angular distribution of the $\kk$ system in the low-invariant-mass region and quantifies the $CP$ asymmetry and branching fraction as a function of the $\kk$ invariant mass and the branching fraction as a function of the $\kpi$ invariant mass. We employ a reoptimized binning close to the kinematic boundaries of the $\kk$ and $\kpi$ systems. This paper also includes the full description of the analysis of the branching fraction and direct $CP$ asymmetry measurement previously published in Ref.~\cite{Hsu}.

The data were collected with the Belle detector at the KEKB asymmetric-energy (3.5 on 8 GeV) 
$e^+e^-$ collider~\cite{kekb}. 
The data sample consists of $772 \times 10^{6}$ $\bb$ pairs accumulated at the $\Upsilon(4S)$ resonance, corresponding to an integrated luminosity of $711~\rm fb^{-1}$, and an additional 
$89~{\rm fb}^{-1}$ of off-resonance data recorded at a center-of-mass (c.m.) energy about $60$ MeV below the $\Upsilon(4S)$ resonance.

\section{\label{sec:belle}Belle Detector and event reconstruction}
The Belle detector consists of a silicon vertex detector (SVD), a 50-layer central drift chamber (CDC), time-of-flight scintillation counters (TOF), an array of aerogel threshold Cherenkov counters (ACC), and a CsI(Tl) electromagnetic calorimeter located inside a superconducting solenoid coil that provides a 1.5~T magnetic field.
Outside the coil, the $K_L^0$ and muon detector, composed of resistive plate counters, detects $K_L^0$ mesons and identifies muons. The detector is described in detail elsewhere~\cite{belled}.
The dataset used in this analysis was collected with two different inner detector configurations. A data sample corresponding to $140~\rm fb^{-1}$ was collected with a beam pipe of radius $2$ cm and with three layers of SVD, while the rest of the data were collected with a beam pipe of radius 1.5 cm and four layers of SVD~\cite{bellesvd2}.
A GEANT3-based~\cite{geant} Monte Carlo (MC) simulation of the Belle detector is used to optimize the event selection and to estimate the signal efficiency.
The signal MC sample is generated with the EvtGen package~\cite{evtgen}, assuming a three-body phase space combined with 
a resonance decaying to two kaons as observed by BaBar and LHCb~\cite{BaBar_kkpi,lhcb_new}.

To reconstruct $\kkpi$, we combine two oppositely charged kaons with a charged pion. 
Charged tracks originating from a $B$ decay are required to have a distance of closest approach with respect to the interaction point of less than 5.0 cm along the $z$ axis (opposite the positron beam direction) and less than 0.2 cm in the $r-\phi$ transverse plane, and a transverse momentum of at least 100 MeV/$c$.

Charged kaons and pions are identified using the charged-hadron identification~(PID) systems, namely CDC, ACC and TOF.
The information from the PID systems is combined to form a $K$-$\pi$ likelihood ratio 
$\mathcal{R}_{K/\pi} = \mathcal{L}_K/(\mathcal{L}_K+\mathcal{L}_\pi)$, 
where ${\cal L}_K$ and ${\cal L}_{\pi}$ are the likelihoods for the kaon and pion hypotheses, respectively. 
Tracks with ${\cal R}_{K/\pi} > 0.6$ are regarded as kaons and those with ${\cal R}_{K/\pi} < 0.4$ as pions. 
With these requirements, the identification efficiencies for 1 GeV/$c$ kaons and pions are $83\%$ and $90\%$, respectively; $6\%$ of 
the pions are misidentified as kaons, and $12\%$ of the kaons are misidentified as pions.

Candidate $B$ mesons are identified using two kinematic variables in the c.m. frame: the beam-energy constrained mass, $\Mbc \equiv \sqrt{\ebeam^{2}/c^{4}-{|p_{B}/c|}^{2}}$, and the energy difference, $\de \equiv E_{B} - \ebeam$, where $E_{B}$ and $p_{B}$ are the reconstructed energy and momentum of $B$ candidates in the c.m. frame, respectively, and $\ebeam$ is the run-dependent beam energy corresponding to half of the c.m. energy. The $B$-meson candidates are required to have $\mb>5.24$~GeV/$c^2$ and $|\de|<0.3$~GeV, and a signal-enhanced region is defined as $5.27 < \mb < 5.29$~GeV/$c^2$ and $|\de|<0.05$~GeV for the optimization of background suppression.
For $19\%$ of the events, there is more than one $B$-meson candidate in an event; we choose the one with the best fit quality from the $B$ vertex fit using the three charged tracks. This criterion selects the correctly reconstructed $B$-meson candidate in $92\%$ of MC events.

\section{\label{sec:bck_std}Background Study}
The dominant background is from continuum $e^+ e^- \to q\overline{q}~(q=u,d,s,c)$ processes. 
The difference on the event topology between the spherical \bb signal and the jetlike continuum background can be exploited; a Fisher discriminant formed from 17 modified Fox-Wolfram moments~\cite{ksfw} is introduced to suppress backgrounds from this source.
To further improve the classification power, we combine the output of the Fisher discriminant with four other discriminating observables in a neural network employed by the NeuroBayes software package~\cite{neubay}. The observables are: the cosine of the angle between the $B$ candidate direction and the beam axis, 
the cosine of the angle between the thrust axis~\cite{thrust} of the $B$ candidate and that of the rest of event 
(both of these quantities being calculated in the c.m. frame), 
the separation along the $z$ axis between the vertex of the $B$ candidate and that of the remaining tracks, 
and the tagging quality variable from a $B$ flavor-tagging algorithm~\cite{btagging}.
The training and optimization of the neural network are performed with signal and continuum MC samples. 
The neural network output ($\nn$) selection requirement is optimized by maximizing a figure of merit 
defined as $N_{\textrm S}/\sqrt{N_{\textrm S}+N_{\textrm B}}$ in the signal-enhanced region defined in the previous paragraph, where $N_{\textrm S}$ denotes 
the expected number of signal events based on MC simulation for a branching fraction of $5\times 10^{-6}$ 
and $N_{\textrm B}$ denotes the expected number of background events. 
The $\nn$ is required to be greater than 0.88, which removes $99\%$ of the continuum events while retaining $48\%$ of the signal.

Background contributions from $B$ decays via the dominant $b \to c$ transition (generic $B$ decays) are investigated with a MC 
sample of such decays. The resulting $\de$ distribution is found to peak strongly in the signal region. 
Peaks are observed in the $\kk$ and $\kpi$ invariant-mass spectra, arising from $b \to c$ decays. These peaking contributions are from $D^{0} \to \kk$ or $K^- \pi^+$
at the nominal $D^0$ mass or from $\chi_{c0}\to \kk$ at the nominal $\chi_{c0}$ mass. In the $\kk$ mass distribution, $D^0\to K^- \pi^+$ form the peak slightly shifted from the $D^0$ mass in the $\mkk$ spectrum owing to $K-\pi$ misidentification. 
To suppress these backgrounds, the candidates for which the invariant mass of the $\kk$ or $\kpi$ system lies in range of $1850-1880$ MeV/$c^2$ are removed. 
The veto region corresponds to $\pm 3.75\sigma$ around the nominal $D^0$ mass, where $\sigma$ is the mass resolution. 
In the case of $K-\pi$ misidentification, we use the pion-mass hypothesis for one of the kaons. For a $\kk$ candidate pair, both $K^+ \pi^{-}$ and $\pi^+K^-$ assignments are tested, and when at least one of them lies within the veto region, the candidate is rejected. A veto mass range of $3375-3475$ MeV/$c^2$ is introduced to reject the backgrounds from $\chi_{c}^{0}\to \kk$ decays.
To suppress the possible charmonium backgrounds from $J/\psi \to \ell^+ \ell^- (\ell=e, \mu)$ decays, 
we apply the electron- or muon-mass hypothesis for both charged daughters, and exclude candidates that lie in the range of $3060-3140$ MeV/$c^2$, 
which corresponds to $\pm 4\sigma$ around the nominal $J/\psi$ mass. Since no significant peak is found in the $\psi(2S)$ mass region, we do not apply a veto for it.

The charmless $B$ decays are studied with a large MC sample where one of the $B$ mesons decays to a charmless final state.
There are a few modes that contribute in the $\mb$ signal region with a corresponding $\de$ peak, denoted collectively as the ``rare peaking''
background. 
\textcolor{black}{These backgrounds are caused by the $K-\pi$ misidentification that leads to the $\de$ peaks shifted toward the negative~(positive) values for $B^+\to \kk K^+$~($B^+\to K^+ \pi^- \pi^+$) and its intermediate resonant modes.}
The remaining contribution other than the peaking components is called the ``rare combinatorial'' background.

\section{\label{sec:fitting}Signal Extraction}
The signal yield is extracted by a two-dimensional extended unbinned maximum likelihood fit in $\mb$ and $\de$ in each $\mkk$ bin, with the likelihood defined as
\begin{equation}
\mathcal{L}=\frac{e^{-\sum_jN_j}}{N!}\prod^{N}_{i}(\sum_j N_j\mathcal{P}^{i}_{j})\mbox{,}
\end{equation}
where
\begin{equation}
\mathcal{P}^{i}_{j}=\frac{1}{2}(1-q^{i}\cdot \mathcal{A}_{CP,j}\xspace) \times \mathcal{P}_{j}(\mb^i,\de^i)\mbox{,}
\end{equation}
\begin{equation}
\textcolor{black}{\mathcal{A}_{CP}=\frac{N_{B^{-}}-N_{B^{+}}}{N_{B^{-}}+N_{B^{+}}}\mbox{.}}
\end{equation}

Here, $N$ is the total number of candidate events, $i$ is the event index, and $N_j$ is the yield of events for category $j$, 
which indexes the signal, continuum, generic $B$, and rare $B$ components. 
$\mathcal{P}_{j}(\mb,\de)$ is the probability density function (PDF) in $\mb$ and $\de$ for the $j$-th
 category. 
The electric charge of the $B$-meson candidate in event $i$ is denoted $q^i$ and $\mathcal{A}_{CP,j}$ is the direct $CP$ asymmetry for category $j$. 
In the signal $B$ decays, there are two cases: all final-state particles are correctly combined (``true'' signal), or one of the daughter particles is a product of the other $B$ decay (``self-cross-feed'' [SCF] contribution). \textcolor{black}{The ratios of SCF contribution to the signal events are found to be between $0.1\% \sim 7.6\%$ in different $\mkk$ bins.}
We prepare the corresponding PDFs, ${\cal P}_{\rm sig}$ and ${\cal P}_{\rm SCF}$. In the fit, the ratio of SCF contribution to the signal events is fixed according to the MC expectation. The signal yield, $\Nsig$, is the yield of the signal PDF.
The signal PDF is represented by the product of a double Gaussian in $\mb$ and a triple Gaussian in $\de$, where the shape parameters are determined from the signal MC sample and 
are calibrated for the possible data-MC difference using a control sample of $B^+ \to \bar{D}^0 \pi^+$ with $\bar{D}^0 \to K^+K^-$ (which we denote as $\control$ hereinafter).
The PDF that describes the continuum background is the product of an ARGUS function~\cite{argus} in $\mb$ and a second-order polynomial in $\de$.
The parameters of the continuum PDF are derived from MC simulation, which agrees with the off-resonance data.
The others~(generic $B$, rare combinatorial, rare peaking, and SCF) are modelled with two-dimensional smoothed histograms from MC simulation due to the strong correlation between $\de$ and $\mb$.
The free parameters in the fit are the signal yield, the signal $\acp$, the generic $B$ yield, the rare peaking yields, and the continuum yield. 
The yields of rare combinatorial backgrounds are also derived from the MC study. 
The $\acp$ of all backgrounds is fixed to zero in the fit.
The stability and bias of the two-dimensional fit is checked by ensemble tests with a set of large-statistics MC events.
The validity of the fit and branching fraction extraction method is checked using data in a high-statistics control sample of the $\control$ decays.
The measured branching fraction for the control sample is consistent with the world-average value~\cite{pdg2016}.

\begin{figure}
 \subfigure[$B^- \to K^- K^+ \pi^-$]{\includegraphics[width=0.24\textwidth]{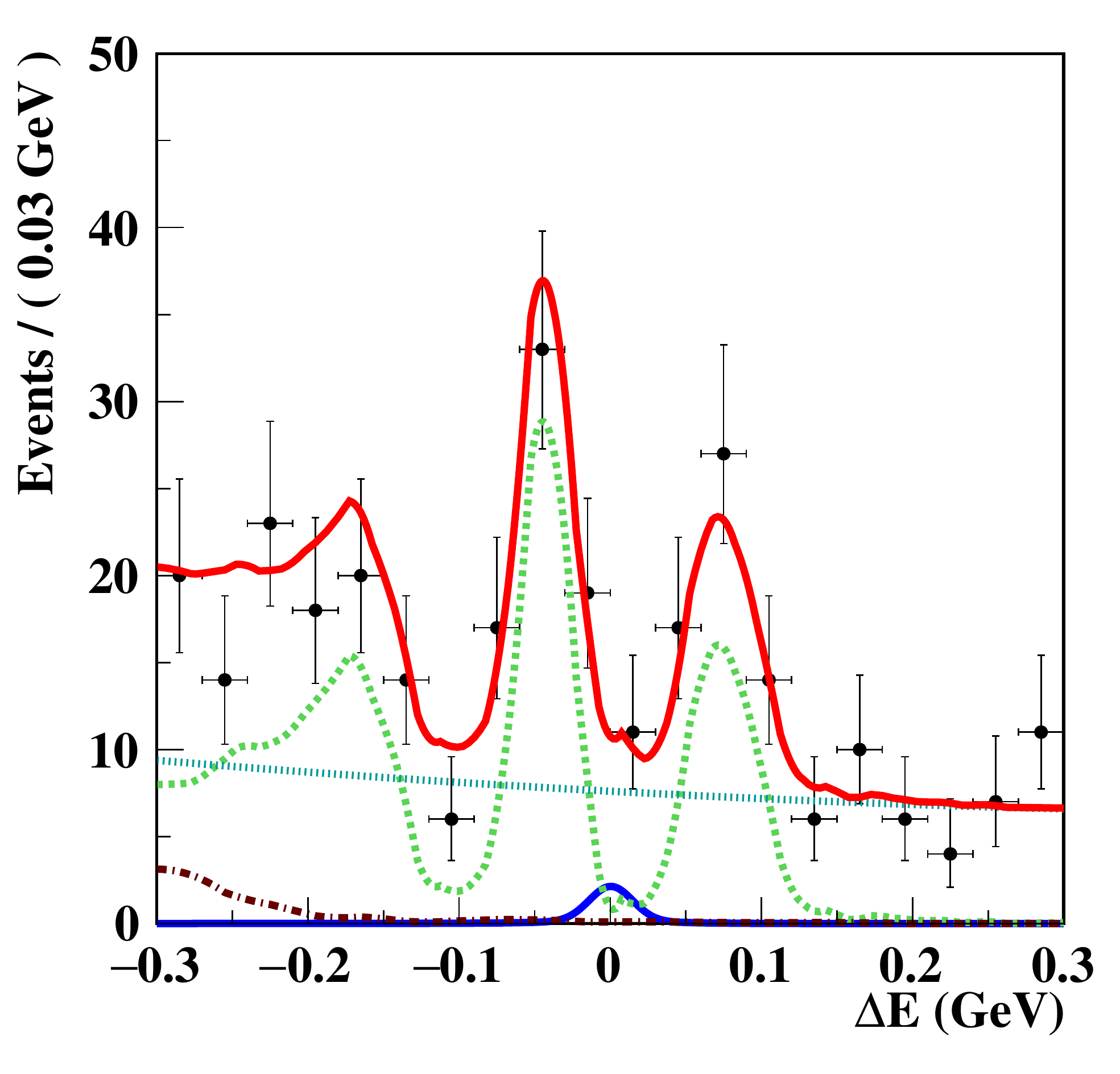}\includegraphics[width=0.24\textwidth]{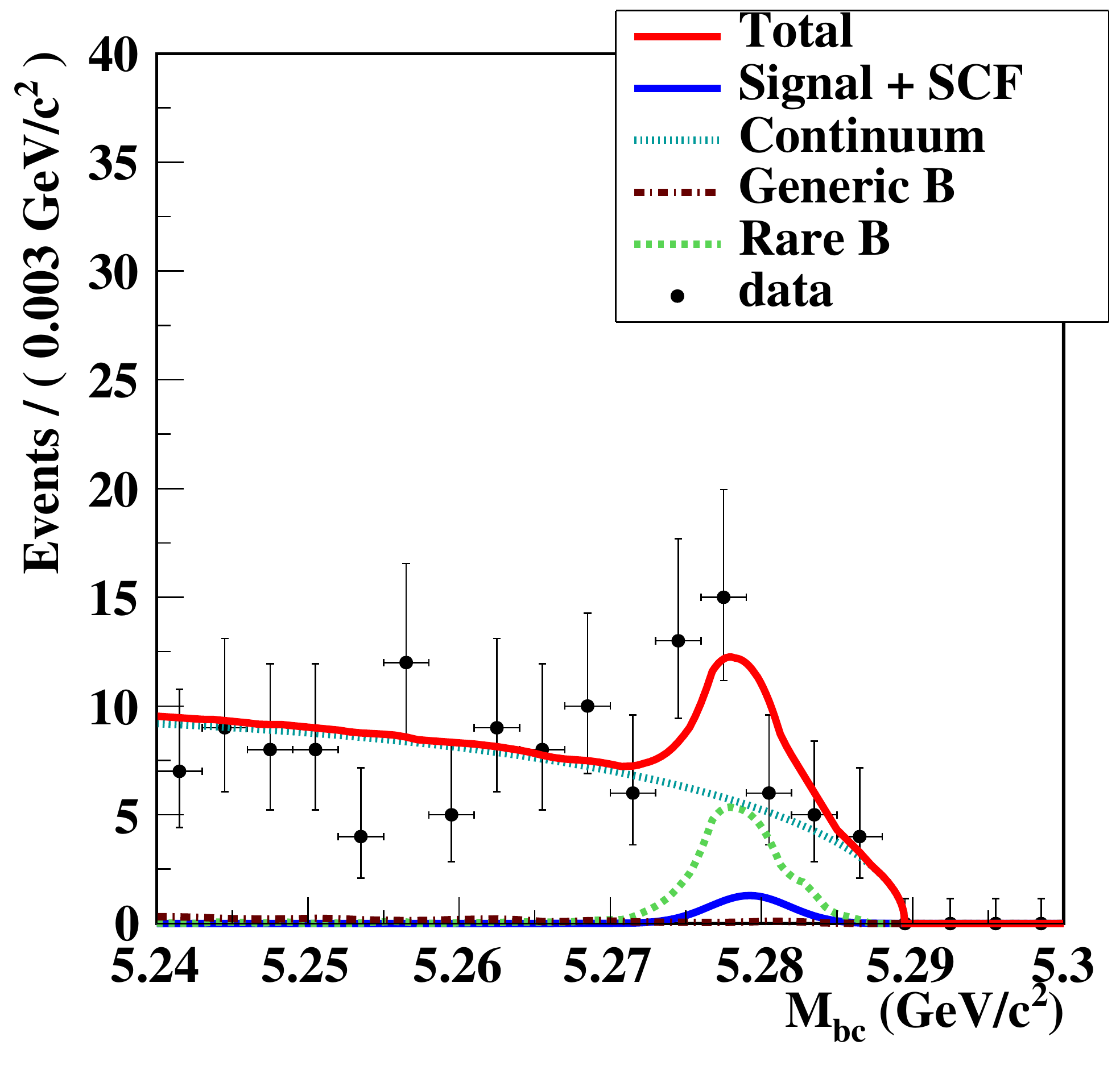}}
 \subfigure[$\kkpi$]{\includegraphics[width=0.24\textwidth]{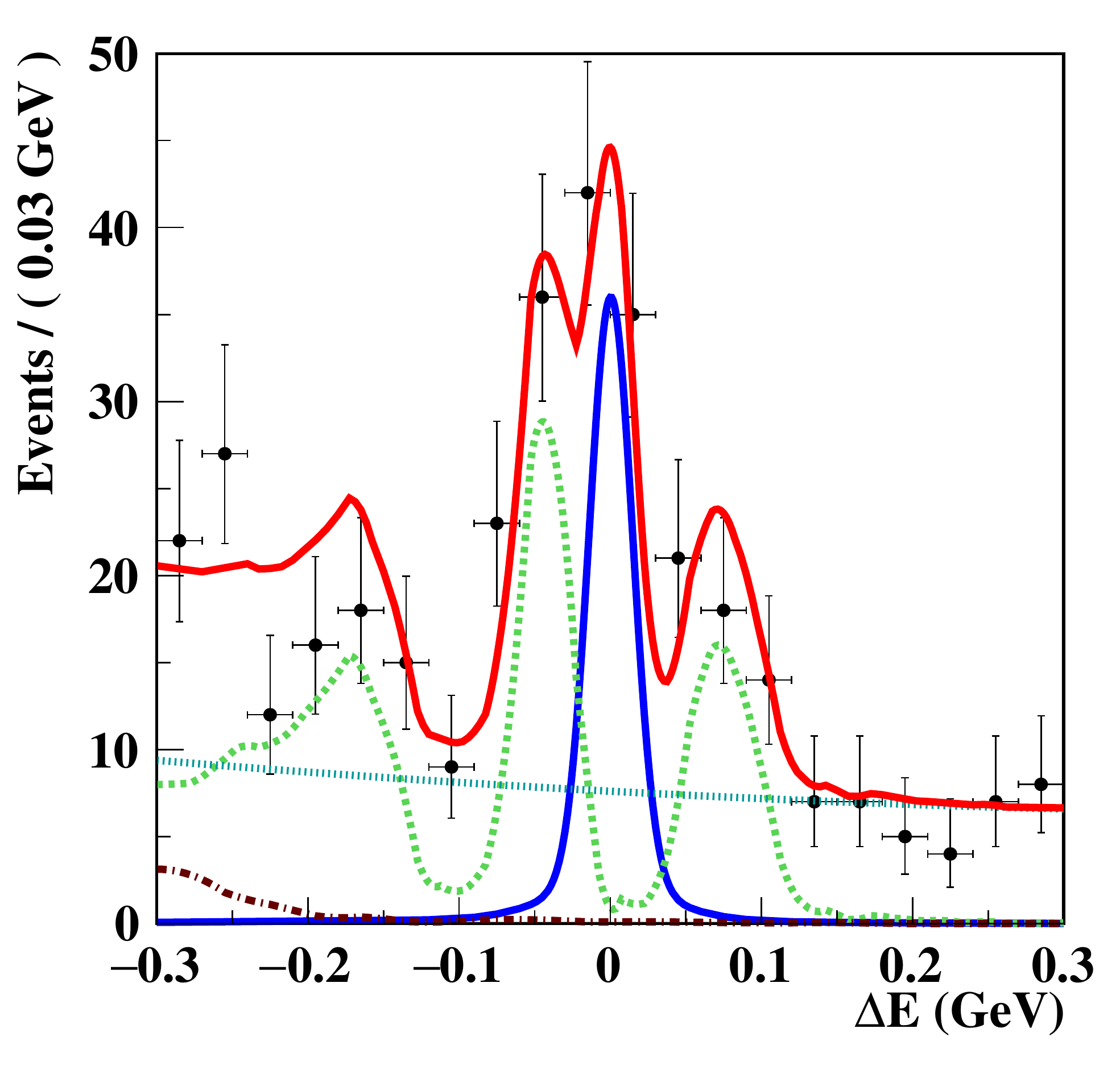}\includegraphics[width=0.24\textwidth]{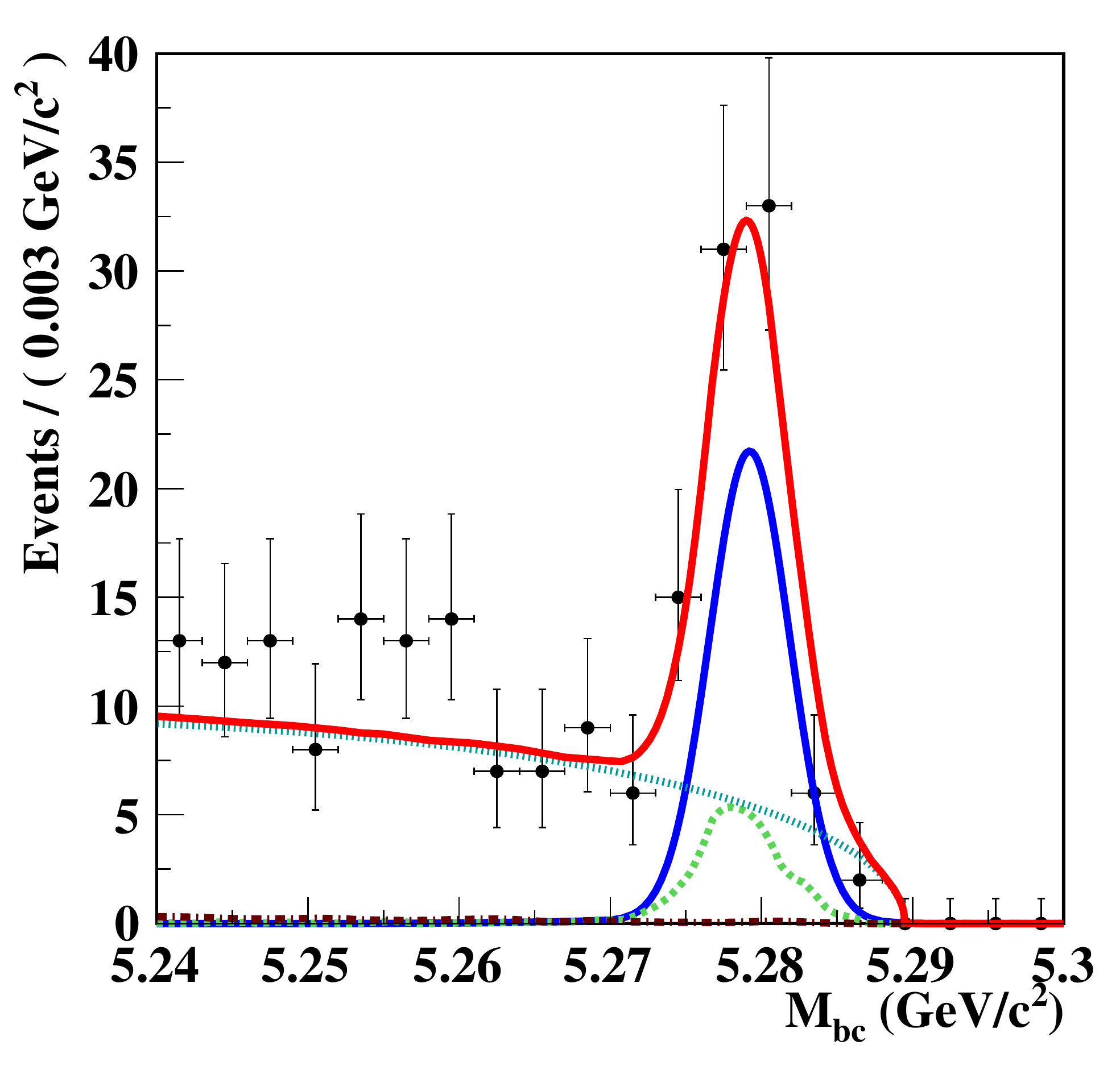}}
\caption{The projections of the $\mb$-$\de$ fit to the data within the range of $0.98 < \mkk < 1.1$ GeV/$c^2$. 
Points with error bars are the data, the red line is the fit result, the blue line is the sum of the signal and the self-cross-feed, 
the cyan dotted line is the continuum background, the brown dot-dashed line is the generic $B$ backgrounds, 
and the green dashed line is the sum of the rare $B$ backgrounds. The projection on $\de$ is with the requirement of $5.275 < \mb < 5.2835$ GeV/$c^2$, while the projection on $\mb$ is with the requirement of $-0.03 < \de < 0.03$ GeV.
}
\label{fig:result_all}
\end{figure}

\begin{figure}
\includegraphics[width=0.9\columnwidth]{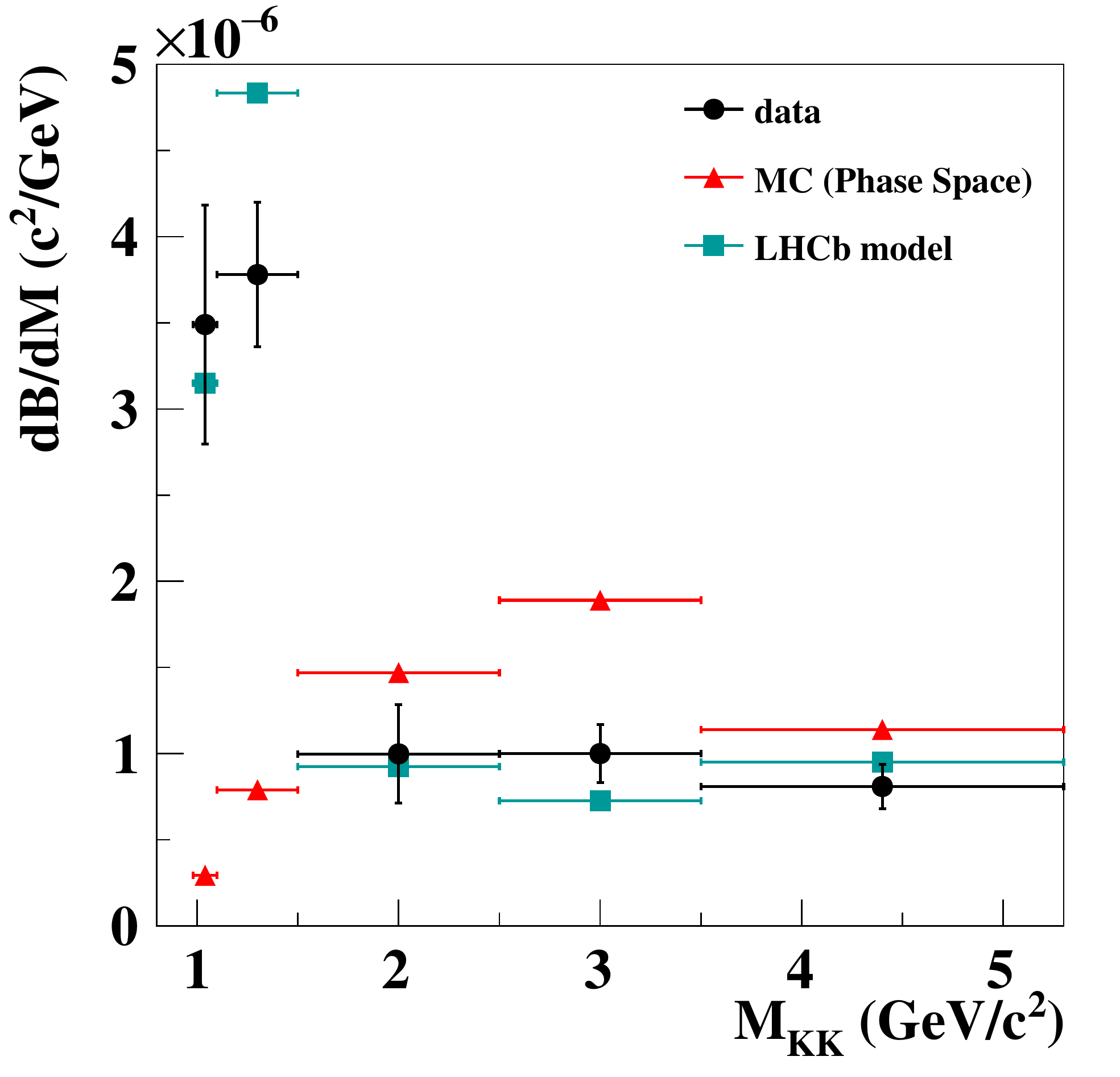}\\
\includegraphics[width=0.9\columnwidth]{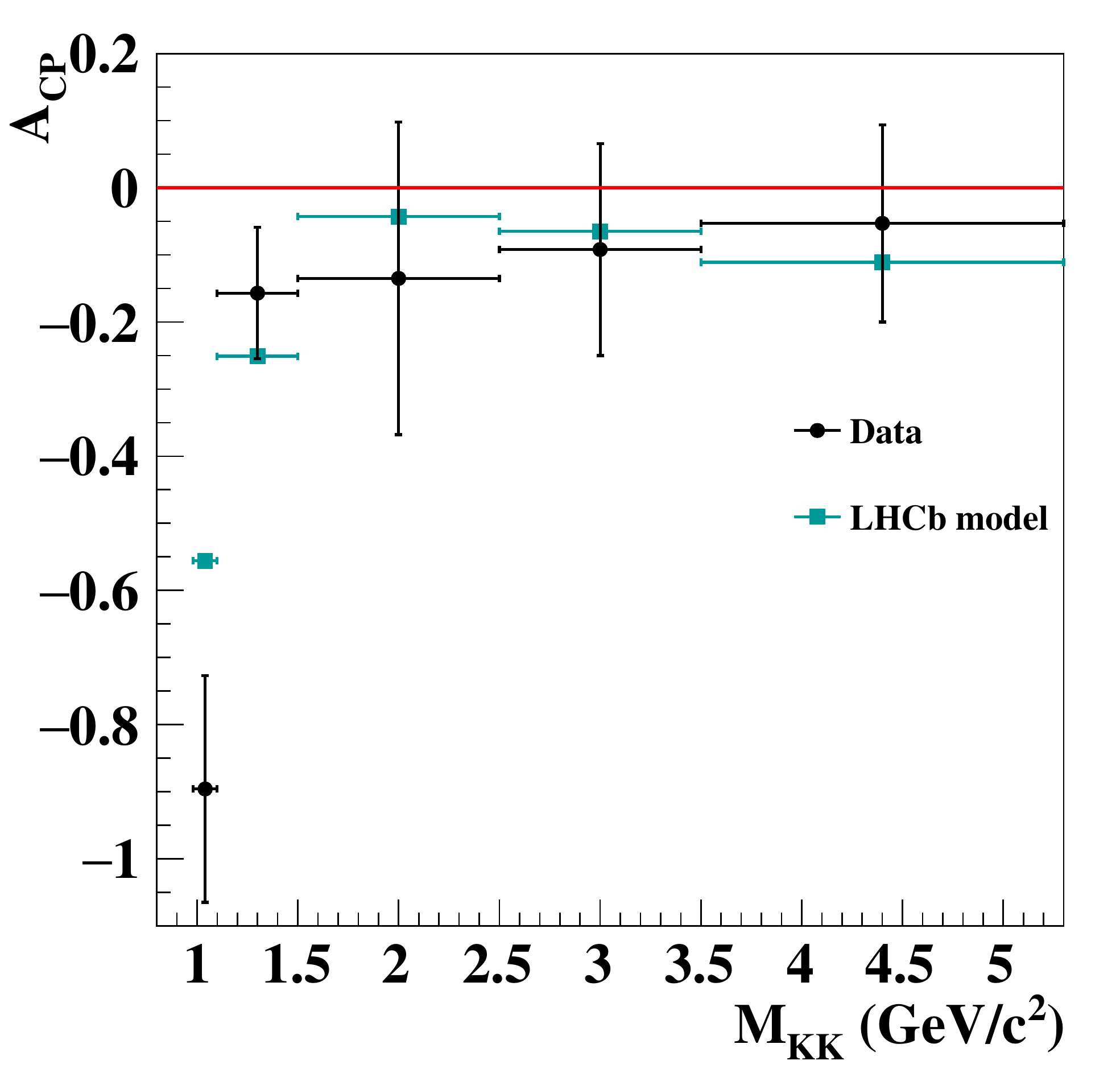}
\caption{
Differential branching fractions~(top) and measured $\acp$~(bottom) as a function of $\mkk$. Each point is obtained from a two-dimensional fit with systematic uncertainty included. Red triangles with error bars in the top figure show the expected signal distribution in a three-body phase space MC, and the cyan squares show the LHCb model which reproduces the model reported by the LHCb~\cite{lhcb_dalitz}. Note that the MC hypotheses/models are rescaled to the experimentally observed total $\kkpi$ signal yield.}
\label{fig:binfit}
\end{figure}

As an example, Fig.~\ref{fig:result_all} shows the fit results within the range of $0.98<\mkk< 1.1$ GeV/$c^2$ in a signal-enhanced region. 
We use the efficiency and fitted yield in each $\mkk$ bin to calculate the branching fraction:
\begin{equation}
\label{eq:bf}
\br=\frac{\Nsig}{\epsilon \times \cpid \times N_{\bb}}\mbox{,}
\end{equation}
where $N_{\bb}$, $\epsilon$, and $\cpid$, respectively, are the number of $\bb$ pairs ($772 \times 10^{6}$), the reconstruction efficiency, 
and the correction factor for particle identification ($94.2\%$) that accounts for the data-MC difference.
We assume that charged and neutral $\bb$ pairs are produced equally at the $\Upsilon(4S)$ resonance.
Table~\ref{tab:binfit} lists the fitted yields, the efficiencies, and the measured $\acp$ in all $\mkk$ bins. Figures of the fit results in all $\mkk$ bins are attached in Appendix~\ref{app:mkk}.
The significance of our measurements is evaluated using the convolution of the likelihood function with a Gaussian function of width equal to the additive systematic uncertainties that only affect the signal yield and $\acp$. Detailed study of the systematic uncertainties is described in Sec.~\ref{sec:syst}. The corresponding significance is given by $\sqrt{-2\ln(\calL_0/\calL_{\rm max})}$,
where $\calL_{\rm max}$ and $\calL_0$ are the likelihood values with and without the signal component, respectively.
Figure~\ref{fig:binfit} shows the results of differential branching fraction and $\acp$ as a function of $\mkk$, where an excess and a large $\acp$ are seen in $\mkk < 1.5$ GeV/$c^{2}$, confirming the observations by BaBar and LHCb.
Our measurements show good agreement with the LHCb model in $\mkk$.
We find strong evidence of a large $CP$ asymmetry of 
$-0.90\pm0.17\pm0.03$ with $4.8\sigma$ significance for $\mkk < 1.1$ GeV/$c^2$.
We integrate the differential branching fractions over the entire mass range of the $\kk$ system to obtain the charge-averaged branching fraction:
\begin{equation}
\br(\kkpi)=(5.38 \pm 0.40\pm 0.35)\times 10^{-6}\mbox{,}
\end{equation}
where the quoted uncertainties are statistical and systematic, respectively.
The weighted average $\acp$ over the entire $\mkk$ region is
\begin{equation}
\acp = -0.170 \pm 0.073 \pm 0.017\mbox{,}
\end{equation}
where the $\acp$ in each $\mkk$ bin is weighted by the fitted yield divided by the detection efficiency in that bin.
The statistical uncertainties are independent among bins; thus, the term is a quadratic sum. 
For the systematic uncertainties, the contribution from the bin-by-bin varying sources is a linear sum while the contribution from
the common sources is a quadratic sum.

\begin{table*}
\caption{Signal yield, efficiency, differential branching fraction, and $\acp$ for individual $\mkk$ bins. 
The first uncertainties are statistical, and the second are systematic. The differential branching fraction is obtained by the partial branching fraction divided by width of that bin.}
\label{tab:binfit}
\begin{center}
\begin{tabular}{c|cccc}
\hline\hline
$\mkk$  & \multirow{2}{*}{$\Nsig$} & \multirow{2}{*}{Eff. ($\%$)} & $\textcolor{black}{d}\BR/\textcolor{black}{d}M$ & \multirow{2}{*}{$\acp$} \\
(GeV/$c^2$) & & & ($\times 10^{-7}$~$c^2$/GeV) \\
\hline
0.98--1.1 & $59.8 \pm 11.4 \pm 2.6 $ & $19.7$ & $34.9 \pm 6.6 \pm 2.0$ & $-0.90 \pm 0.17 \pm 0.04$ \\
1.1--1.5 & $212.4 \pm 21.3 \pm 6.7$ & $19.3$ & $37.8 \pm 3.8 \pm 1.9$ & $-0.16 \pm 0.10 \pm 0.01$ \\
1.5--2.5 & $113.5 \pm 26.7 \pm 18.6$ & $15.6$ & $10.0 \pm 2.3 \pm 1.7$ & $-0.15 \pm 0.23 \pm 0.03$ \\
2.5--3.5 & $110.1 \pm 17.6 \pm 4.9$ & $15.1$ & $10.0 \pm 1.6 \pm 0.6$ & $-0.09 \pm 0.16 \pm 0.01$ \\
3.5--5.3 & $172.6 \pm 25.7 \pm 7.4$ & $16.3$ & $8.1 \pm 1.2 \pm 0.5$ & $-0.05 \pm 0.15 \pm 0.01$ \\
\hline\hline
\end{tabular}
\end{center}
\end{table*}

\section{\label{sec:heli}Angular distribution of the \kk system in the low \mkk region}
From the fit result in the lowest $\mkk$ bin, we obtain only 3.1 signal events in the $B^{-}$ sample and 56.7 events in the $B^{+}$ sample within the signal enhanced region. To determine the spin configuration of the \kk system for the signal in $\mkk<1.1$~GeV/$c^{2}$, the signal yields are obtained by fitting to the data samples in bins of the helicity angle\textcolor{black}{, and are summarized in Table~\ref{tab:costheta}}. The helicity angle, $\theta_{\rm hel}$, is defined as the angle between the momenta of the $B$ meson and the kaon with the same charge as the $B$ meson, as both are evaluated in the \kk rest frame. Due to the large $\acp$ observed in the data sample, only the $B^+$ candidates are used in the fit.

\begin{table}
\textcolor{black}{
\caption{Signal yields and efficiency for individual $\cos\theta_{\rm hel}$ bins.}
\label{tab:costheta}
\begin{center}
\begin{tabular}{c|cc}
\hline\hline
$\cos\theta_{\rm hel}$  & $\Nsig$ & Eff.(\%) \\
\hline
$-1.0 - -0.6$ & $7.2 \pm 3.7 \pm 0.7$ & 18.4 \\
$-0.6 - -0.2$ & $7.0 \pm 3.8 \pm 0.5$ & 19.7 \\
$-0.2 - 0.2$ & $13.3 \pm 4.6 \pm 0.7$ & 21.6 \\
$0.2 - 0.6$ & $11.8 \pm 4.4 \pm 0.5$ & 21.2 \\
$0.6 - 1.0$ & $14.7 \pm 5.2 \pm 0.7$ & 20.2 \\
\hline\hline
\end{tabular}
\end{center}
}
\end{table}

\begin{figure}
\includegraphics[width=0.9\columnwidth]{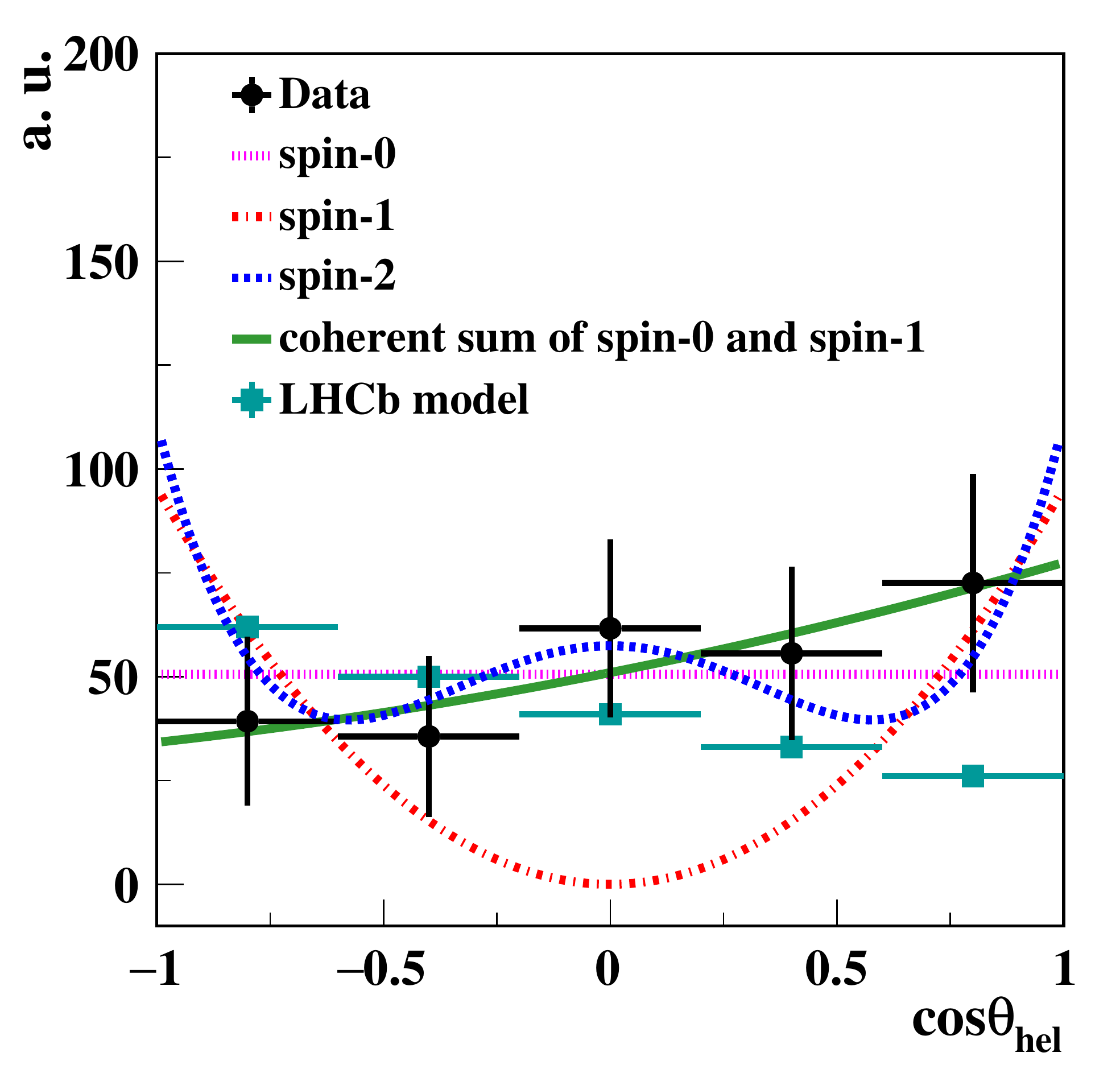}
\caption{The helicity angle distribution with efficiency correction applied, in arbitrary units. The fits using spin-0, spin-1, and spin-2 models and the coherent sum of spin 0 and and spin 1 are superimposed. The LHCb model was calculated according to Ref.~\cite{lhcb_dalitz}.}
\label{fig:coshel_fit}
\end{figure}

Since there is no evidence for the narrow $\phi$ state in the $\mkk$ distribution and there is no other resonance between $0.98~{\rm GeV/}c{^2} < \mkk < 1.1~{\rm GeV/}c{^2}$, $a~priori$ one would expect no specific spin configuration in the data. 
To investigate this, we generated MC samples with the ansatz of $B^{+} \to X_{\kk}\pi^{+}$ with $X_{\kk} \to \kk$ using different assumptions for the spin state of $X$. \textcolor{black}{Here $X$ is described by a relativistic Breit-Wigner function that is centered at $1.2$ GeV/$c^{2}$ with a width of $0.25$ GeV/$c^{2}$\textcolor{black}{, which is estimated based on Ref.~\cite{lhcb_old}}.}
The fit results using different models for the helicity distribution are shown in Fig.~\ref{fig:coshel_fit}. With the application of the efficiency correction, the spin-0 component is described by a constant, while the spin-1 and -2 components are described by functions $a\times |P_{1}|^2$ and $a(1+b\times |P_{2}|^2)$, respectively. Here $P_{N}$ is an $N$th-order Legendre polynomial.

Besides the models of pure $S$-wave, $P$-wave, and $D$-wave, we also test with a coherent-sum model of the $S$-wave and $P$-wave,
\begin{equation}
\begin{aligned}
    f(\cos\theta_{\rm hel}) = A_{S}^2 &+ 2A_{S}A_{P}\cos\theta_{SP} P_{1}(\cos\theta_{\rm hel})\\
    &+A_{P}^2P^{2}_{1}(\cos\theta_{\rm hel})\mbox{,}
\end{aligned}
\end{equation}
and obtain the magnitudes of the amplitudes from the $S$-wave and $P$-wave to be $A_{S}=7.1\pm 1.0$ and $A_{P}=2.2 \pm 8.6$ in arbitrary units; the ratio between these two is
\begin{equation}
r=\frac{A_{P}}{A_{S}}=0.31 \pm 1.23 \mbox{.}
\end{equation}
At the $1\sigma$ level, the data are consistent with the vanishing $P$-wave amplitude; therefore, the relative phase is very poorly constrained: $\cos\theta_{SP} = 0.69 \pm 2.58$. 
The forward-backward asymmetry is
\begin{equation}
    A_{FB} = \frac{A_{S}A_{P}\cos\theta_{SP}}{A_{S}^2 + \frac{A_{P}^2}{3}}=0.21 \pm 0.17\mbox{.}
\end{equation}
The $p$-value of each model is summarized in Table~\ref{tab:model}. As the reduced $\chi^2$~\textcolor{black}{($\chi^2/ndf$)} for pure $P$-wave is much larger than unity, the distribution is unlikely to be described by the pure $P$-wave. The LHCb model is not inconsistent with our data but less favoured than the spin-0, spin-2, and coherent-sum models.

\begin{table}
\caption{Fit results for different angular distribution of $X_{\kk}$. The $p$-value corresponds to the \textcolor{black}{ reduced $\chi^{2}$}. The results show that, except the coherent sum and LHCb model, other components are unlikely.}
\label{tab:model}
\begin{center}
\begin{tabular}{c|ccccc}
\hline\hline
  & $X_{J=0}$ & $X_{J=1}$ & $X_{J=2}$ & Coherent sum & LHCb model \\
\hline
$\chi^{2}/ndf$ & 1.9/4 & 14.4/4 & 1.6/3 & 0.5/2 & 7.0/4 \\
$p$-value & 0.750 & 0.006 & 0.815 & 0.792 & 0.136 \\
\hline\hline
\end{tabular}
\end{center}
\end{table}

\section{\label{sec:mkpi}Differential Branching fraction as a function of $\mkpi$}
\begin{figure}[h]
\includegraphics[width=0.48\columnwidth]{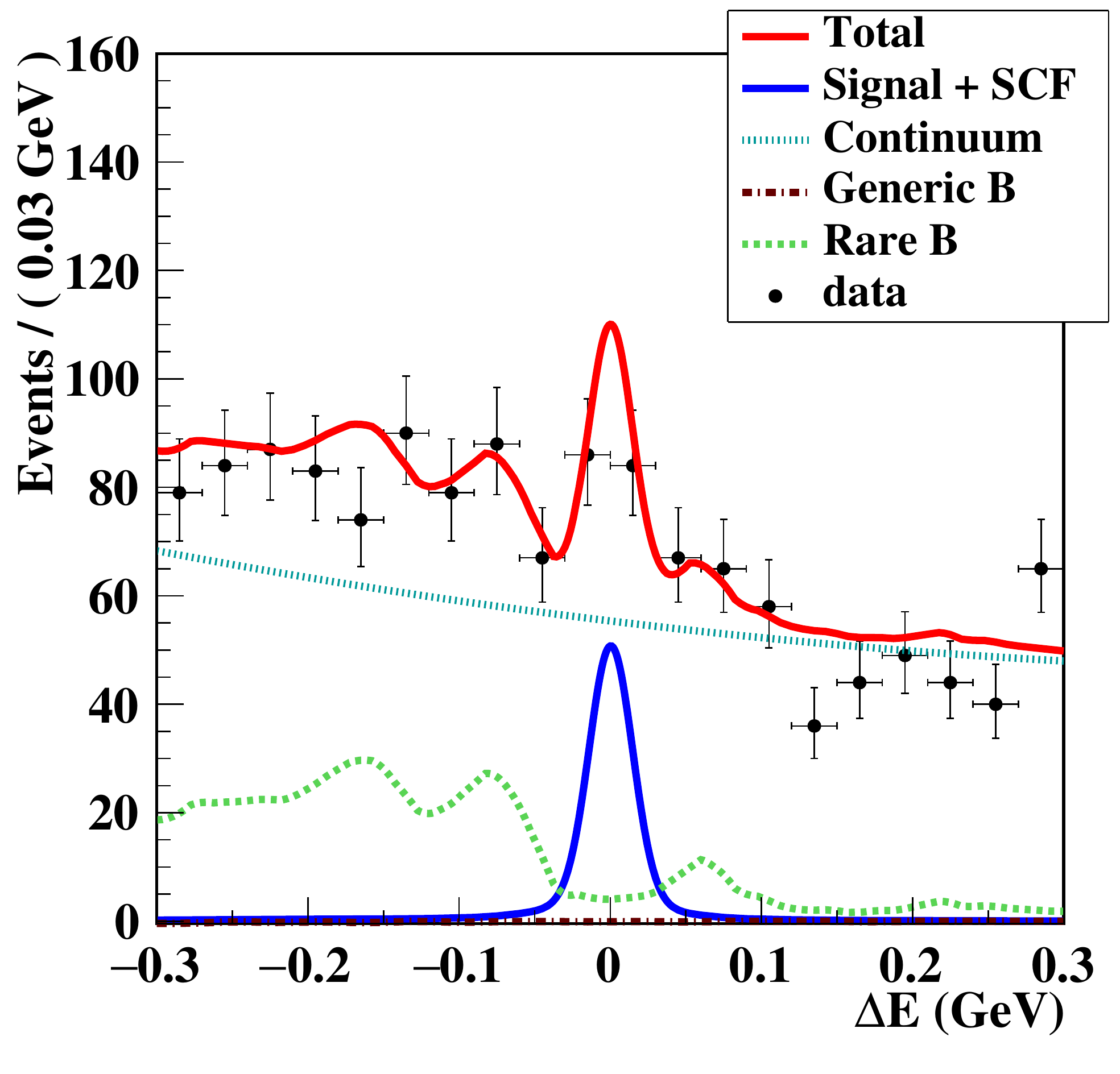}
\includegraphics[width=0.48\columnwidth]{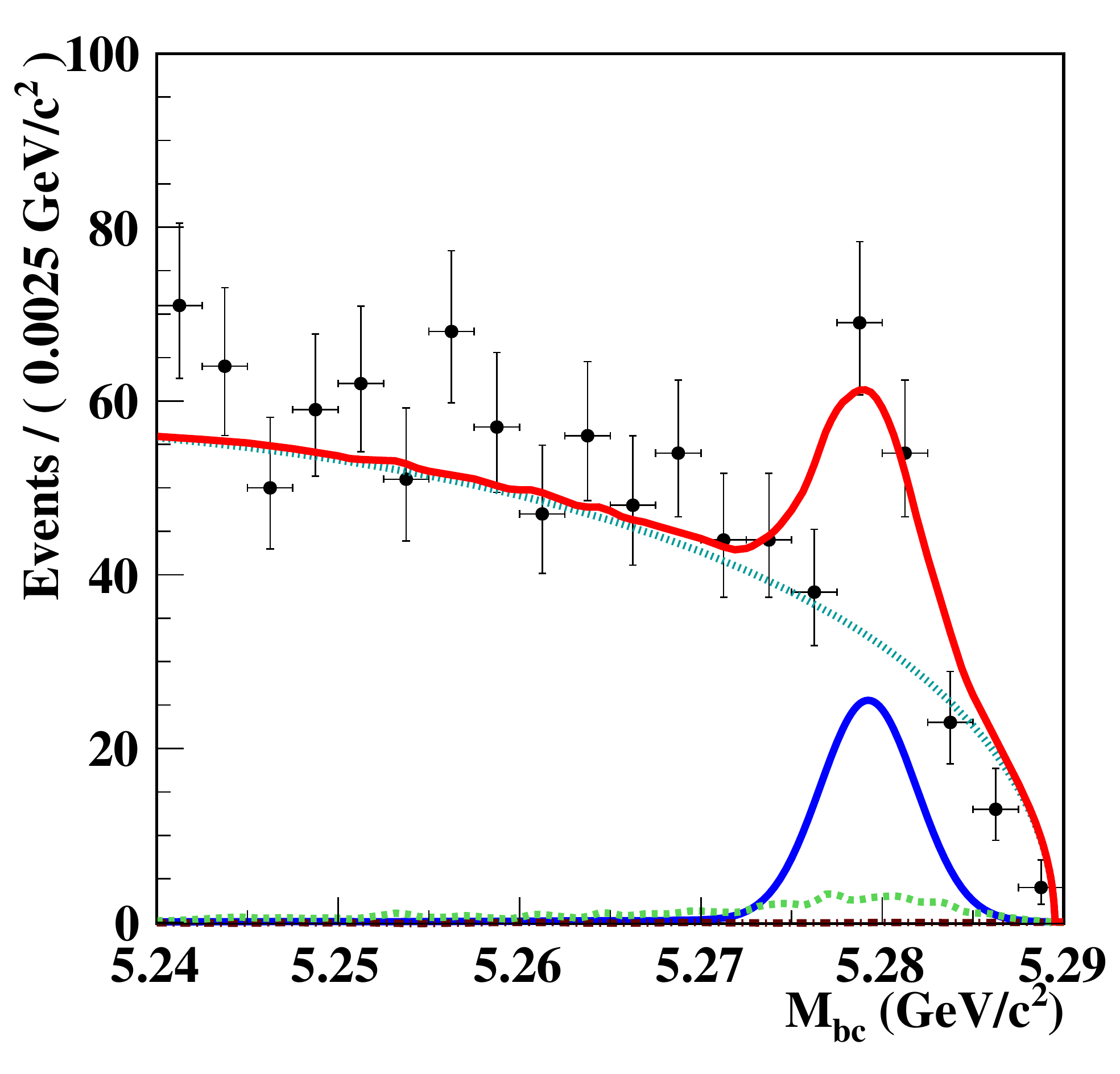}
\caption{Signal-enhanced projections of the $\mb$-$\de$ fit to data in the first $\mkpi$ bin.   
Points with error bars are the data, the red line is the fit result, the blue line is the sum of the signal and the self cross feed, 
the cyan dotted line is the continuum background, the brown dot-dashed line is the generic $B$ backgrounds, 
and the green dashed line is the sum of the rare $B$ backgrounds. The projection on $\de$ is with the requirement of $5.275 < \mb < 5.2835$~GeV/$c^2$, while the projection on $\mb$ is with the requirement of $-0.03<\de<0.03$~GeV.}
\label{fig:result_mkpibin}
\end{figure}

\begin{figure}
\includegraphics[width=0.9\columnwidth]{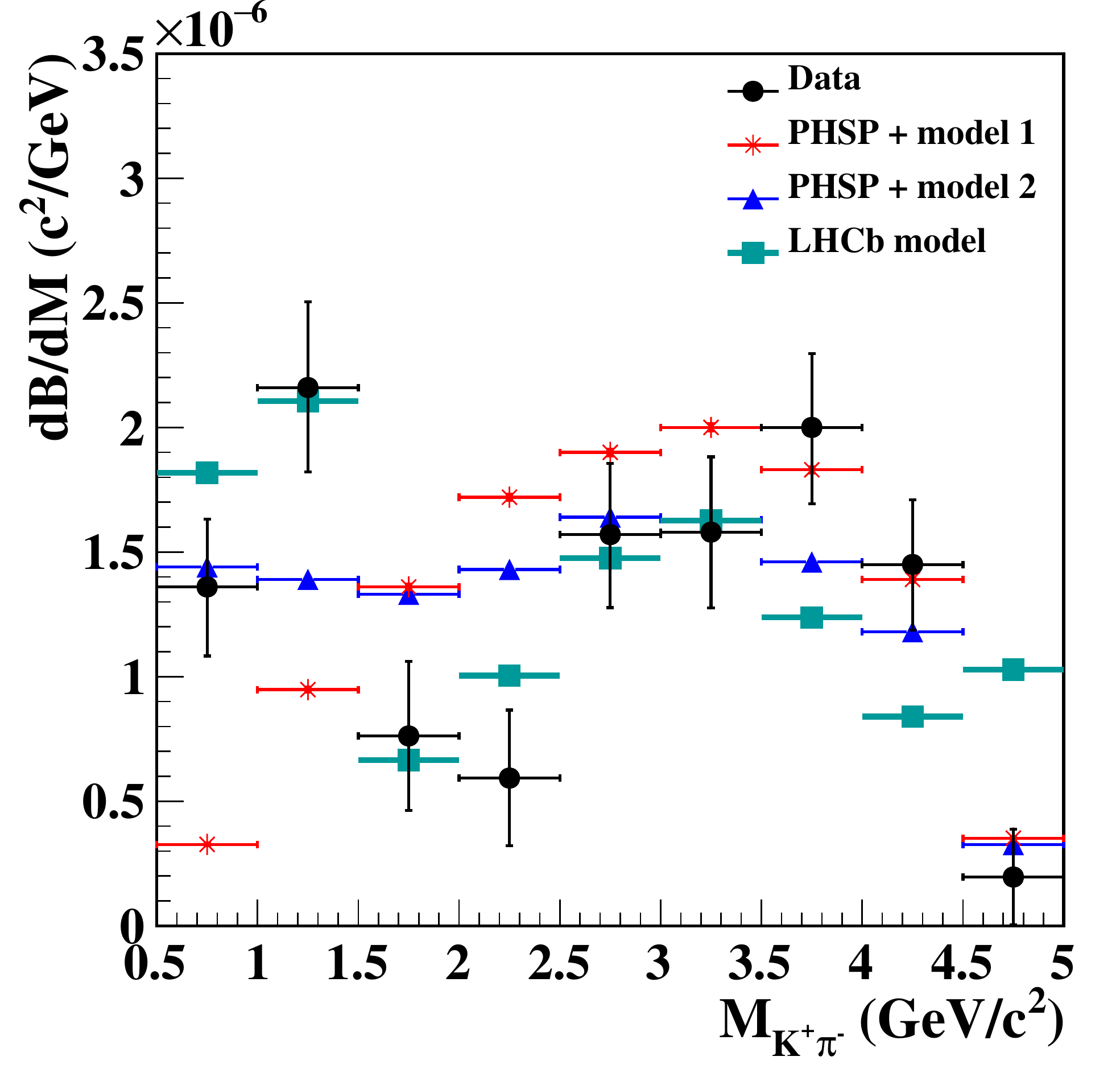}
\caption{Differential branching fractions as a function of $\mkpi$. Each point is obtained from a two-dimensional fit with systematic uncertainty included. Also shown are two ansatzes as red asterisks and blue triangles. The red asterisks show 3-body phase space combined with a 10\% contribution from $B^{+}$ decays to $X_{\kk}\pi^{+}$ with spin-0, noted as model 1 in the figure. The blue triangles show the effect of adding the expected contribution of the intermediate states, $B^{+}\to K^{*0}K^{+}$ and $B^{+}\to K^{*0}_{0}K^{+}$, noted as model 2.
The cyan squares show the model which reproduces the model proposed from Ref.~\cite{lhcb_dalitz}.
Note that the MC hypotheses/models are rescaled to the experimentally observed total signal yield.}
\label{fig:mkpicom}
\end{figure}

To study the distribution and also to provide additional information in the Dalitz plane of $\kkpi$, we also provide the measurement of the differential branching fraction in $\mkpi$. The analysis procedure is exactly the same as described in Sec.~\ref{sec:fitting}, except with the data sample now divided into 9 bins in the range of $\mkpi$ from 0.5 GeV/$c^{2}$ to 5.0 GeV/$c^{2}$. No $\acp$ was measured from the fit. Table~\ref{tab:mkpibinfit} lists the fitted yields and the differential branching fractions for all \textcolor{black}{$\mkpi$} bins.
As an example, Fig.~\ref{fig:result_mkpibin} shows the fit results of the first $\mkpi$ bin in a signal-enhanced region. The fit results of all $\mkpi$ bins can be found in Appendix~\ref{app:mkpi}.
\textcolor{black}{Both} the LHCb result~\cite{lhcb_new} and the differential branching fraction shown in Fig.~\ref{fig:binfit} indicate that the resonant contributions for the $K^+K^-$ \textcolor{black}{projection} are estimated to be 10\% of the total $B^+ \to K^+ K^- \pi^+$ signal\textcolor{black}{. As} a possible approximation to describe it, the $X_{K^+K^-}$ with spin 0 is introduced (model 1). For the remaining \textcolor{black}{90\% of $B^{+}$ events in the $\kk$ dimension, and for all $B^{-}$ events}, the hypothesis to \textcolor{black}{only} have \textcolor{black}{a} three-body phase space decay is disfavored. Instead, \textcolor{black}{another hypothesis that takes \textcolor{black}{$B^{+}\to \overline{K}^{*}(892)^{0}K^{+}$} and \textcolor{black}{$B^{+}\to \overline{K}^{*}_{0}(1430)^{0}K^{+}$} components into account according to the current world average in Ref.~\cite{pdg2020}\textcolor{black}{~(11\% and 6\% of the inclusive \kkpi branching fraction, respectively)}, and considers the three-body phase space decay to account for the remaining signal yield, is proposed. T}his hypothesis (model 2) \textcolor{black}{is} found to be compatible when it comes to explaining the obtained differential branching fraction as a function of $\mkpi$, as shown in Fig.~\ref{fig:mkpicom}. For models 1 and 2, the reduced $\chi^2$ values are found to be 6.5 and 2.9, respectively. Here, there are eight degrees of freedom.
A direct comparison with the LHCb model is also made; despite the good agreement in $\mkpi<\textcolor{black}{3.5}$ GeV/$c^2$, the reduced $\chi^2$ of this model against our measurements is 4.5, which is lower than that of model 1.

\renewcommand{\arraystretch}{1.5}
\begin{table}
\caption{Signal yield and differential branching fraction for each $\mkpi$ bin. 
The first uncertainties are statistical and the second are systematic.}
\label{tab:mkpibinfit}
\begin{center}
\begin{tabular}{c|cccc}
\hline\hline
$\mkpi$  & \multirow{2}{*}{$N_{sig}$} & \multirow{2}{*}{Eff.~(\%)} & $\textcolor{black}{d}\BR/\textcolor{black}{d}M$ \\
(GeV/$c^2$) & &  & ($\times 10^{-7}$~$c^2$/GeV)\\
\hline
0.5--1.0 & $78.7 \pm 15.3 \pm 3.7 $ & 16.0 & $13.6 \pm 2.6 \pm 0.6$ \\
1.0--1.5 & $128.6 \pm 19.6 \pm 6.1 $ & 16.4 & $21.6 \pm 3.3 \pm 1.0$ \\
1.5--2.0 & $41.4 \pm 15.9 \pm 3.7 $ & 14.9 & $ 7.6 \pm 2.9 \pm 0.7$ \\
2.0--2.5 & $34.1 \pm 14.7 \pm 5.5 $ & 15.8 & $5.9 \pm 2.6 \pm 1.0$ \\
2.5--3.0 & $93.3 \pm 16.3 \pm 4.4 $ & 16.3 & $15.7 \pm 2.8 \pm 0.7$ \\
3.0--3.5 & $77.5 \pm 14.4 \pm 3.4 $ & 13.5 & $15.8 \pm 2.9 \pm 0.7$ \\
3.5--4.0 & $126.3 \pm 17.7 \pm 6.0 $ & 17.4 & $20.0 \pm 2.8 \pm 1.0$ \\
4.0--4.5 & $100.9 \pm 17.2 \pm 5.3 $ & 19.2 & $14.5 \pm 2.5 \pm 0.8$ \\
4.5--5.0 & $12.4 \pm 11.5 \pm 4.1$ & 17.5 & $1.9 \pm 1.8 \pm 0.6$ \\
\hline\hline
\end{tabular}
\end{center}
\end{table}

\section{\label{sec:syst}Systematic uncertainty}
There are two types of systematic uncertainties: bin-by-bin and bin-independent. The former type varies in different $\mkk$ bins, i.e., background PDF modelling, fixed yields in the fit and background $\acp$. The latter includes sources from the total number of $\bb$ pairs in the Belle data, tracking efficiency, particle identification, continuum suppression, signal PDF modelling, and the possible bias of the fit.
Systematic uncertainties in the partial branching fraction in $\mkk$ bins are itemized in Table~\ref{tab:sysm}, while those for the measurements in $\mkpi$ are listed in Table~\ref{tab:sysm_mkpi}.
The uncertainty due to the total number of $\bb$ pairs is $1.4\%$. The uncertainty due to the charged-track reconstruction efficiency is estimated to be $0.35\%$ per track by using partially reconstructed $D^{*+} \to D^{0}\pi^{+}$, $D^{0}\to \pi^{+}\pi^{-}\ks$ events. 
The uncertainty due to the $\rkpi$ requirements is determined by a control sample study of $D^{*+}\to D^{0}\pi^{+}_{s}$ with $D^{0}\to K^{-}\pi^{+}$, where $\pi^{+}_{s}$ denotes a low-momentum charged pion. This low-momentum pion allows the event to be reconstructed with good signal/noise ratio without relying on particle identification information. Since the particle identification efficiency is related to the momenta of the tracks, the efficiency and fake rate were studied in different momentum and polar angle bins. In the end, we obtained an efficiency correction of 94.2\% and a corresponding uncertainty of 1.4\%.

The uncertainties due to the continuum suppression and the signal PDF shape are estimated using a control sample of $\control$ decays. The potential fit bias is evaluated by performing an ensemble test comprising $1000$ pseudoexperiments, where the signal component is taken from the signal MC sample, and the backgrounds are generated using the shapes of their PDFs obtained from the MC samples.
The observed $2.3\%$ bias is included in the systematic uncertainty calculation.
The uncertainty due to the continuum background PDF modeling is evaluated by varying the PDF  
parameters by $\pm 1\sigma$ of their statistical errors.
The uncertainty due to the data/MC difference is taken into account by using the fit model from the off-resonance data, which is included in the background PDF modeling in Table~\ref{tab:sysm}.
For the $\bb$ background PDFs that are modeled by two-dimensional 
smoothed histogram PDFs, the associated uncertainty is evaluated by changing the bin sizes. 
The uncertainty due to the fixed yields of rare combinatorial backgrounds is also evaluated by varying each fixed yield up or down by its statistical error. 
The uncertainty due to nonzero $\acp$ of rare peaking backgrounds is estimated by assuming the LHCb measurements and their uncertainties~\cite{lhcb_new}.
In the absence of knowledge of the distribution of the SCF component in $\mkk$ and $\mkpi$ bins, we use a conservative approach to evaluate the uncertainty by varying the fraction by $\pm 50\%$; the resulting deviation from the nominal value is included in the fixed yields in Tables~\ref{tab:sysm} and~\ref{tab:sysm_mkpi}.

The $\acp$ systematic errors due to the fixed yields, background $\acp$, and the background PDF modeling are estimated with the same procedure as applied for the branching fraction. 
A possible detector bias due to tracking acceptance and $\rkpi$ is evaluated using the measured asymmetry from the off-resonance data, $\acp^\textrm{off}$\textcolor{black}{. $\acp^\textrm{off}$ is defined as
\begin{equation}
\begin{aligned}
\acp^\textrm{off} &= \acp^\textrm{off, real} + \acp^\textrm{det} \\ &=\frac{N^\textrm{\textcolor{black}{off}}_{K^+K^-\pi^-}-N^\textrm{\textcolor{black}{off}}_{K^+K^-\pi^+}}{N^\textrm{\textcolor{black}{off}}_{K^+K^-\pi^-}+N^\textrm{\textcolor{black}{off}}_{K^+K^-\pi^+}}\mbox{,}
\end{aligned}
\end{equation}
where $\acp^\textrm{off, real}$ represents true $CP$ asymmetry in the dataset, and $\acp^\textrm{det}$ represents the contribution from detector effects. Because $\acp^\textrm{off, real}$ is predicted to be zero in the off-resonance data, the measured $\acp^\textrm{off}$ equals the $\acp^\textrm{det}$.}
We introduce the efficiency-corrected number of $K^+K^-\pi^\pm$ candidates~, $T_{K^+K^-\pi^\pm}$, as:
\begin{equation}
T_{K^+K^-\pi^\pm} = \frac{N^\textrm{\textcolor{black}{off}}_{K^+K^-\pi^\pm}}{\epsilon(K^+)\cdot\epsilon(K^-)\cdot\epsilon(\pi^\pm)}\mbox{,}
\end{equation}
where $N^\textrm{\textcolor{black}{off}}_{K^+K^-\pi^\pm}$ is the observed number of $K^+K^-\pi^\pm$ candidates in the off-resonance data and $\epsilon(h)$ is the detection efficiency of hadron $h$.
Due to the $K^{+} K^{-}$ pair appearing in both $B^{+}$ and $B^{-}$ candidates, the efficiencies they contribute cancel at the first order to give
\begin{equation}
\acp^\textrm{off} = \frac{\epsilon(\pi^{-})T_{K^+K^-\pi^{-}}-\epsilon(\pi^{+})T_{K^+K^-\pi^{+}}}
{\epsilon(\pi^{-})T_{K^+K^-\pi^{-}}+\epsilon(\pi^{+})T_{K^+K^-\pi^{+}}}\mbox{.}
\end{equation}
The asymmetry of the detector efficiency between particles and antiparticles depends on the momentum and polar angle of these daughters in the laboratory frame. In this analysis, $K^{+}$ and $K^{-}$ in the decay product have similar distributions. In addition to this, since we have a pair of $K^{+}$ and $K^{-}$ in the final state, the detector bias for kaons will cancel at the first order. To conservatively estimate the remaining potential asymmetry, we reweight the signal MC sample with the following asymmetric $K^+$ and $K^-$ efficiencies:
\begin{equation}
\epsilon(\vec{p})=
\begin{cases}
~0.8+0.04\times|\vec{p}|,\text{~~for }K^{+} \\
~1.0-0.04\times|\vec{p}|,\text{~~for }K^{-}\mbox{,}\\
\end{cases}
\end{equation}
and the total reweighted factor is $\epsilon({K^+})\cdot\epsilon(K^-)$. The original asymmetry of the testing sample is $0.0017 \pm 0.0019~({\rm stat.})$ and that of the reweighted sample is $0.0021 \pm 0.0019~({\rm stat.})$. The conclusion is that the asymmetries from the original and reweighted samples are consistent and hence any asymmetric efficiency due to kaons has a negligible contribution to the systematic uncertainty.
Therefore, for this analysis, we only need to consider the detector asymmetry from charged pions.

We apply the same criteria as those for the signal except for the continuum-suppression requirement in order to increase the statistics in the off-resonance data.
The value of $\acp^\textrm{off}$ is $0.0024 \pm 0.0014$. \textcolor{black}{Because it is consistent with zero within two standard deviations, we do not correct the detector bias for the measured $\acp$.} The central shift plus $1\sigma$ statistical error is quoted as the systematic uncertainty for the detector bias. 
The systematic uncertainties in $\acp$ are included in Table~\ref{tab:sysm}.

\begin{table*}
\caption{\label{tab:sysm}Systematic uncertainties in the measured branching fraction and $\acp$ in the individual $\mkk$ bins. The dagger ($\dagger$) indicates $\mkk$ dependence of the uncertainty. The dash ($-$) indicates a value below $0.1\%$~($0.001$) in $\BR$~(\acp).}
\begin{center}
\begin{tabular}{lccccc}
\hline\hline
{Source} & \multicolumn{5}{c}{Relative uncertainties in ${\BR}$ ($\%$)} \\
$\mkk$(GeV/$c^2$) & 0.98--1.1 & 1.1--1.5 & 1.5--2.5 & 2.5--3.5 & 3.5--5.3 \\
\hline
Number of $\bb$ pairs & \multicolumn{5}{c}{$1.4$} \\
Tracking & \multicolumn{5}{c}{$1.1$} \\
Particle identification & \multicolumn{5}{c}{$1.4$} \\
Continuum suppression & \multicolumn{5}{c}{$1.3$} \\
Signal PDF & \multicolumn{5}{c}{$1.8$} \\
Fit bias & \multicolumn{5}{c}{$2.3$} \\
Background PDF$^\dagger$ & $3.7$ & $2.2$ & $16.2$ & $3.8$ & $3.6$ \\
Fixed yields$^\dagger$ & $-$ & $-$ & $-$ & 0.1 & $-$ \\
Background $\acp$$^\dagger$ & $0.2$ & $0.3$ & $1.5$ & $0.8$ & $0.4$ \\
\hline\hline
Source & \multicolumn{5}{c}{Absolute uncertainties in $\acp$} \\
$\mkk$(GeV/$c^2$) & 0.98-1.1 & 1.1-1.5 & 1.5-2.5 & 2.5-3.5 & 3.5-5.3 \\
\hline
Background PDF$^\dagger$ & 0.036 & 0.005 & 0.028 & 0.006 & 0.003 \\
Fixed yields$^\dagger$ & $-$ & $-$ & $-$ & 0.002 & $-$ \\
Background $\acp$$^\dagger$ & 0.015 & 0.004 & 0.009 & 0.005 & 0.002 \\
Detector bias &  \multicolumn{5}{c}{$0.004$} \\
\hline\hline
\end{tabular}
\end{center}
\end{table*}

\begin{table*}
\caption{\label{tab:sysm_mkpi}Systematic uncertainties in the measured branching fraction in the individual $\mkpi$ bins. The dagger ($\dagger$) indicates 
$\mkpi$ dependence of the uncertainty. The dash ($-$) indicates a value below $0.1\%$ in $\BR$.}
\begin{center}
\begin{tabular}{lccccccccc}
\hline\hline
{Source} & \multicolumn{9}{c}{Relative uncertainties in ${\BR}$ ($\%$)} \\
$\mkpi$(GeV/$c^2$) & 0.5--1.0 & 1.0--1.5 & 1.5--2.0 & 2.0--2.5 & 2.5--3.0 & 3.0--3.5 & 3.5--4.0 & 4.0--4.5 & 4.5--5.0 \\
\hline
Number of $\bb$ pairs & \multicolumn{9}{c}{$1.4$} \\
Tracking & \multicolumn{9}{c}{$1.1$} \\
Particle identification & \multicolumn{9}{c}{$1.4$} \\
Continuum suppression & \multicolumn{9}{c}{$1.3$} \\
Signal PDF & \multicolumn{9}{c}{$1.8$} \\
Fit bias & \multicolumn{9}{c}{$2.3$} \\
Background PDF$^\dagger$ & $2.7$ & $2.7$ & $8.0$ & $15.7$ & $2.7$ & $2.0$ & $2.8$ & $3.5$ & $32.4$ \\
Fixed yields$^\dagger$ & $0.7$ & $0.4$ & $0.4$ & $0.3$ & $-$ & $0.4$ & $0.2$ & $0.1$ & $0.6$ \\
\hline\hline
\end{tabular}
\end{center}
\end{table*}

\section{Conclusion}
In conclusion, we report the measured branching fraction and direct $CP$ asymmetry for the suppressed decay $\kkpi$ using the full $\Upsilon(4S)$ data sample collected with the Belle detector. 
We employ a two-dimensional fit to determine the branching fraction as a function of $\mkk$ and $\mkpi$, and $\acp$ as a function of $\mkk$.
We confirm the excess and nonzero local $\acp$ in the low $\mkk$ region reported by LHCb, and quantify the differential branching fraction in each $\kk$ and $\kpi$ invariant 
mass bin. We find an evidence of $4.8\sigma$ significance for a negative $CP$ asymmetry in the region $\mkk < 1.1$ GeV/$c^{2}$. 
Our measured values of the overall branching fraction and direct $CP$ asymmetry are 
$\BR(\kkpi) = (5.38\pm0.40\pm0.35)\times 10^{-6}$ and $\acp = -0.170\pm0.073\pm0.017$, respectively. 
The measurement challenges the conventional description of direct $CP$ violation~\cite{bellecpv_nature} since it requires large contributions to separate weak tree and strong penguin amplitudes in the same small region of phase space in order to simultaneously enhance both the yield and provide the cancellation required for such a large $CP$ effect. So, for example, if the enhancement were due to a large final-state resonance in a strong penguin diagram, there would have to be an accompanying tree-level process of the same magnitude and opposite phase to provide the almost complete cancellation observed in the measurement.
The Dalitz analysis performed by LHCb finds a significant contribution from $S$-wave $\pi^{+}\pi^{-} \to K^{+}K^{-}$ rescattering, particularly at low invariant mass. Our angular analysis shows a more complex spin structure than the pure $P$-wave in the lowest \mkk bin and the LHCb model is not inconsistent with our data.
Overall, the local enhancement in yield and direct $CP$ asymmetry at low \mkk are well reproduced by the LHCb amplitude model. Understanding this with an {\it ab initio} calculation remains an interesting challenge.

\section{Acknowledgements}
We wish to thank the LHCb Collaboration for providing calculations for the $\mkk$, $\mkpi$ and helicity distributions using their model for $\kkpi$ decays.
We thank the KEKB group for the excellent operation of the accelerator; the KEK cryogenics group for the efficient operation of the solenoid.
This work, based on data collected using the Belle detector, which was
operated until June 2010, was supported by 
the Ministry of Education, Culture, Sports, Science, and
Technology (MEXT) of Japan, the Japan Society for the 
Promotion of Science (JSPS), and the Tau-Lepton Physics 
Research Center of Nagoya University; 
the Australian Research Council including grants
DP180102629, 
DP170102389, 
DP170102204, 
DE220100462, 
DP150103061, 
FT130100303; 
Austrian Federal Ministry of Education, Science and Research (FWF) and
FWF Austrian Science Fund No.~P~31361-N36;
the National Natural Science Foundation of China under Contracts
No.~11675166,  
No.~11705209;  
No.~11975076;  
No.~12135005;  
No.~12175041;  
No.~12161141008; 
Key Research Program of Frontier Sciences, Chinese Academy of Sciences (CAS), Grant No.~QYZDJ-SSW-SLH011; 
Project ZR2022JQ02 supported by Shandong Provincial Natural Science Foundation;
the Ministry of Education, Youth and Sports of the Czech
Republic under Contract No.~LTT17020;
the Czech Science Foundation Grant No. 22-18469S;
Horizon 2020 ERC Advanced Grant No.~884719 and ERC Starting Grant No.~947006 ``InterLeptons'' (European Union);
the Carl Zeiss Foundation, the Deutsche Forschungsgemeinschaft, the
Excellence Cluster Universe, and the VolkswagenStiftung;
the Department of Atomic Energy (Project Identification No. RTI 4002) and the Department of Science and Technology of India; 
the Istituto Nazionale di Fisica Nucleare of Italy; 
National Research Foundation (NRF) of Korea Grant
Nos.~2016R1\-D1A1B\-02012900, 2018R1\-A2B\-3003643,
2018R1\-A6A1A\-06024970, RS\-2022\-00197659,
2019R1\-I1A3A\-01058933, 2021R1\-A6A1A\-03043957,
2021R1\-F1A\-1060423, 2021R1\-F1A\-1064008, 2022R1\-A2C\-1003993;
Radiation Science Research Institute, Foreign Large-size Research Facility Application Supporting project, the Global Science Experimental Data Hub Center of the Korea Institute of Science and Technology Information and KREONET/GLORIAD;
the Polish Ministry of Science and Higher Education and 
the National Science Center;
the Ministry of Science and Higher Education of the Russian Federation, Agreement 14.W03.31.0026, 
and the HSE University Basic Research Program, Moscow; 
University of Tabuk research grants
S-1440-0321, S-0256-1438, and S-0280-1439 (Saudi Arabia);
the Slovenian Research Agency Grant Nos. J1-9124 and P1-0135;
Ikerbasque, Basque Foundation for Science, Spain;
the Swiss National Science Foundation; 
the Ministry of Education and the Ministry of Science and Technology of Taiwan;
and the United States Department of Energy and the National Science Foundation.
These acknowledgements are not to be interpreted as an endorsement of any
statement made by any of our institutes, funding agencies, governments, or
their representatives.
We thank the KEKB group for the excellent operation of the
accelerator; the KEK cryogenics group for the efficient
operation of the solenoid; and the KEK computer group and the Pacific Northwest National
Laboratory (PNNL) Environmental Molecular Sciences Laboratory (EMSL)
computing group for strong computing support; and the National
Institute of Informatics, and Science Information NETwork 6 (SINET6) for
valuable network support.

\onecolumngrid
\clearpage
\appendix
\section{Fit results in \mkk bins\label{app:mkk}}
\begin{figure}[!ht]
\raggedright
\subfigure[$0.8<\mkk<1.1$ GeV/$c^{2}$]{\includegraphics[width=0.24\textwidth]{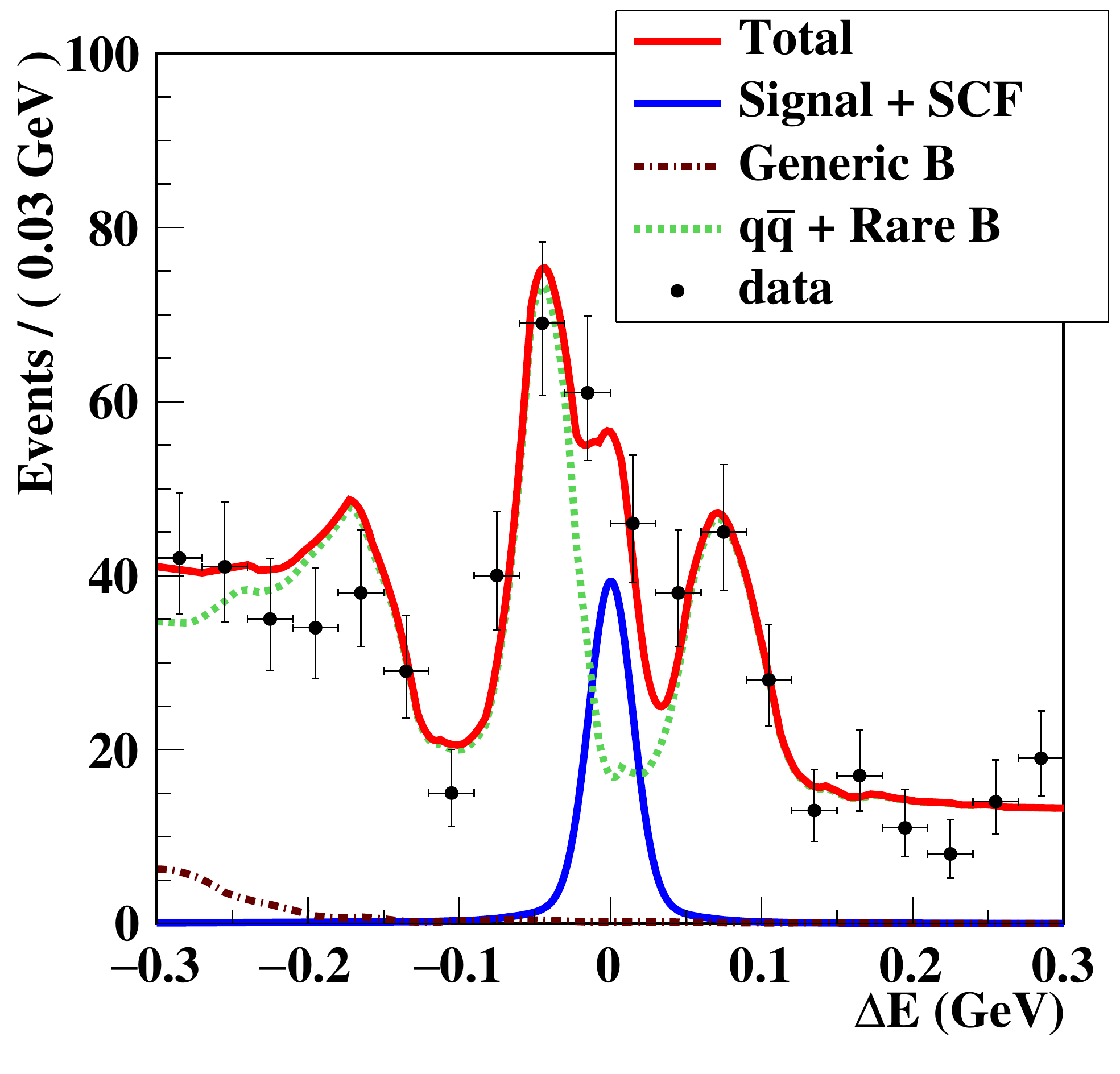}\includegraphics[width=0.24\textwidth]{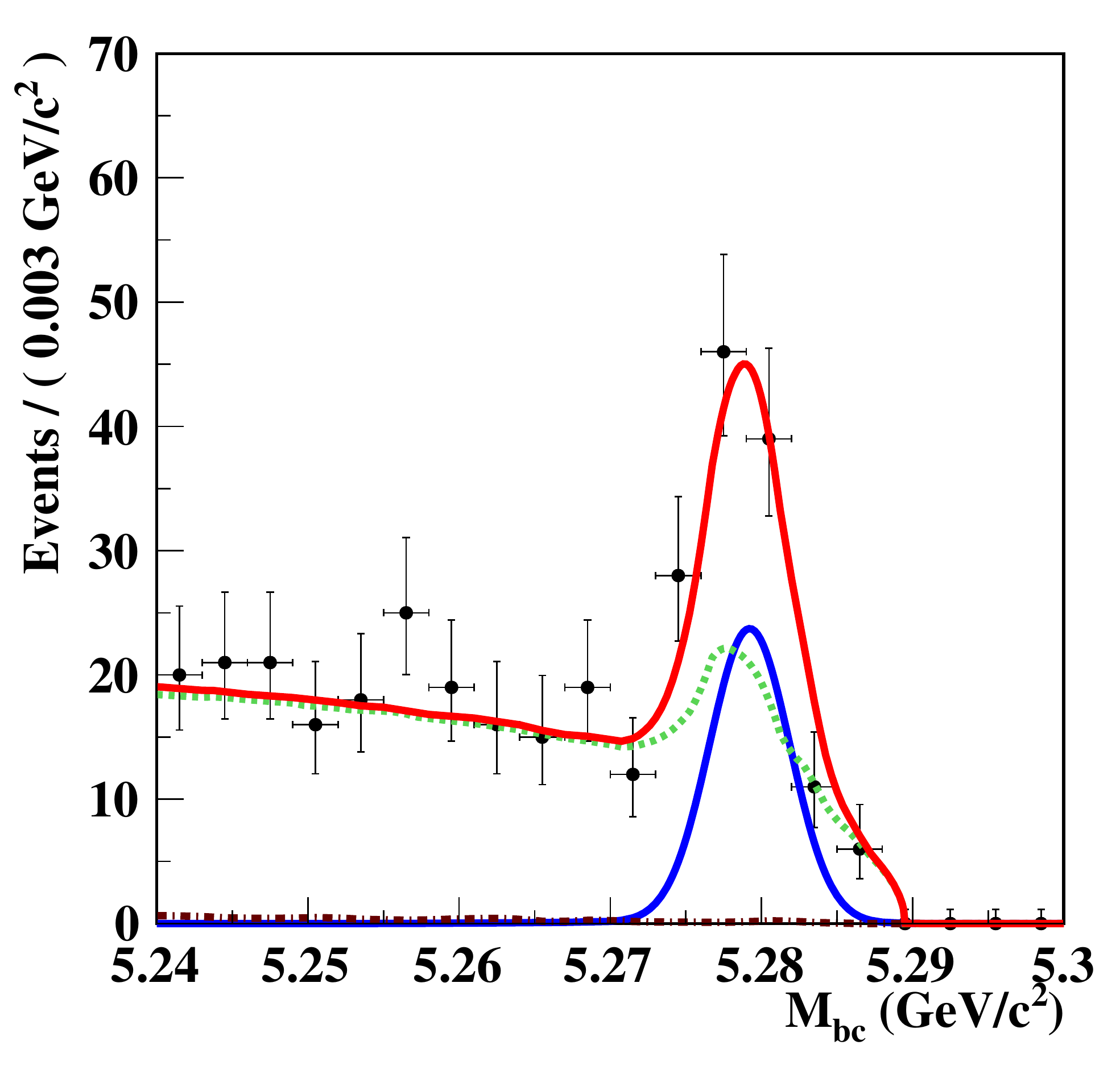}}
\subfigure[$1.1 < \mkk < 1.5$ GeV/$c^2$]{\includegraphics[width=0.24\textwidth]{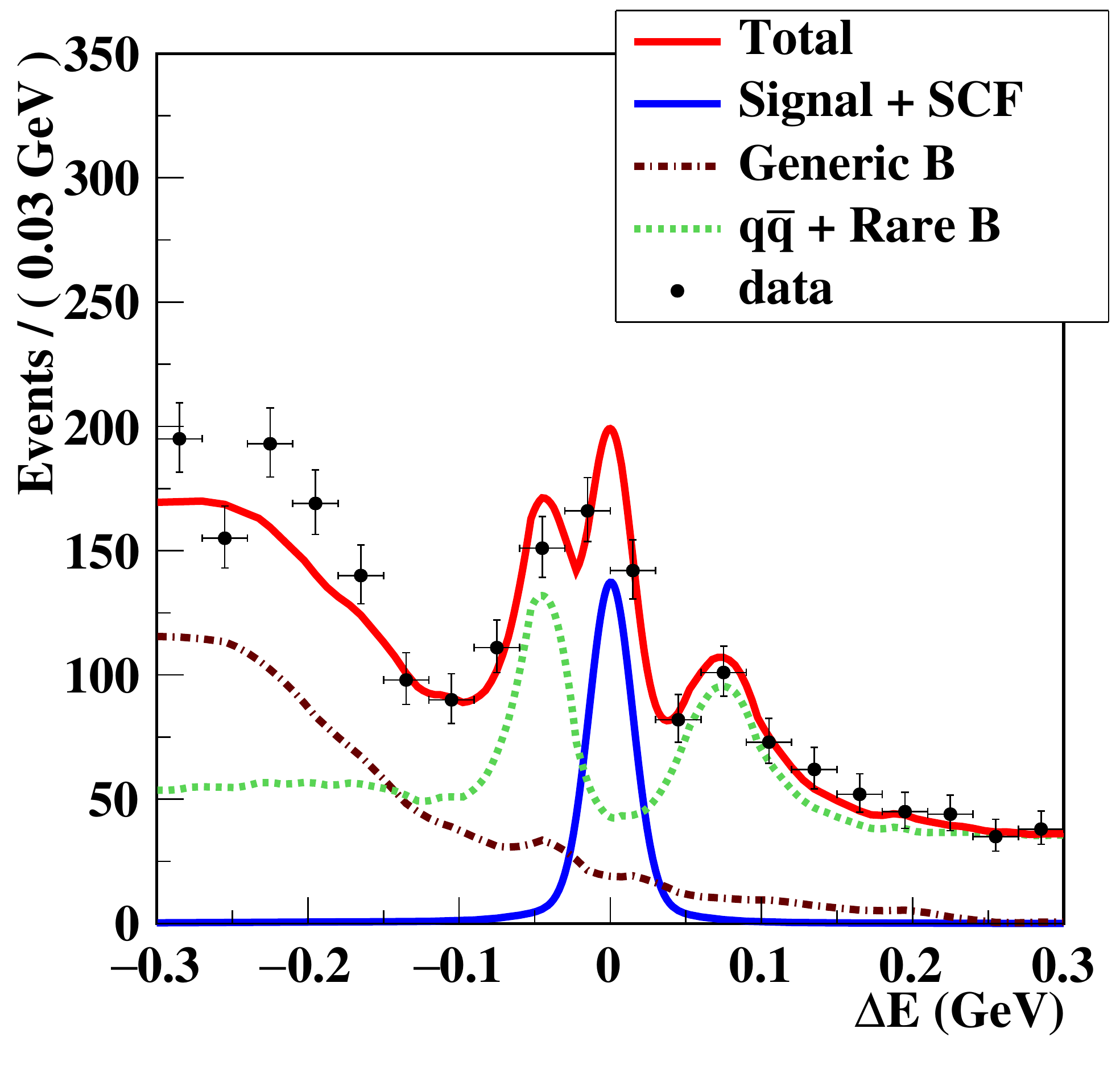}\includegraphics[width=0.24\textwidth]{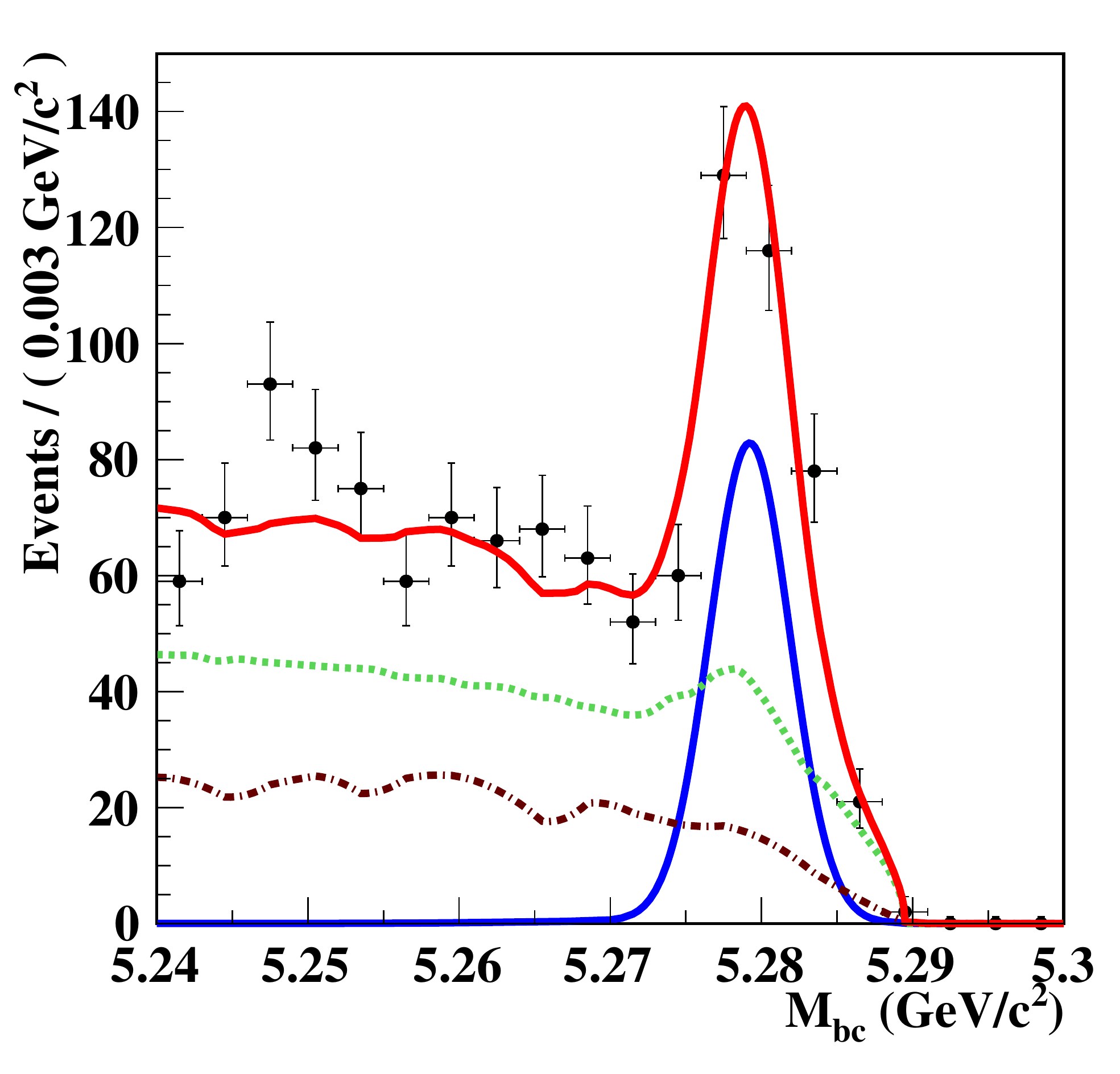}}
\subfigure[$1.5 < \mkk < 2.5$ GeV/$c^2$]{\includegraphics[width=0.24\textwidth]{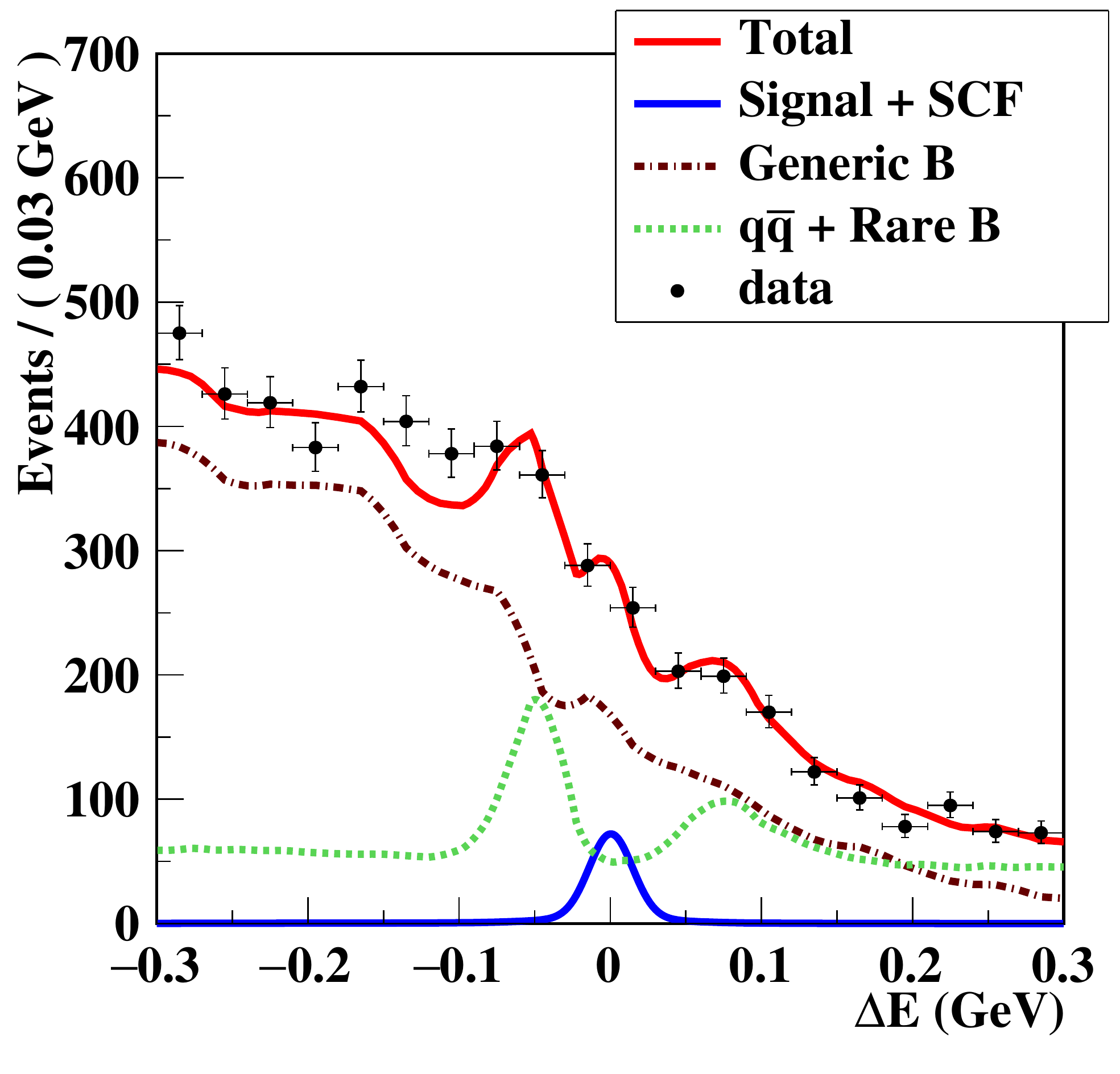}\includegraphics[width=0.24\textwidth]{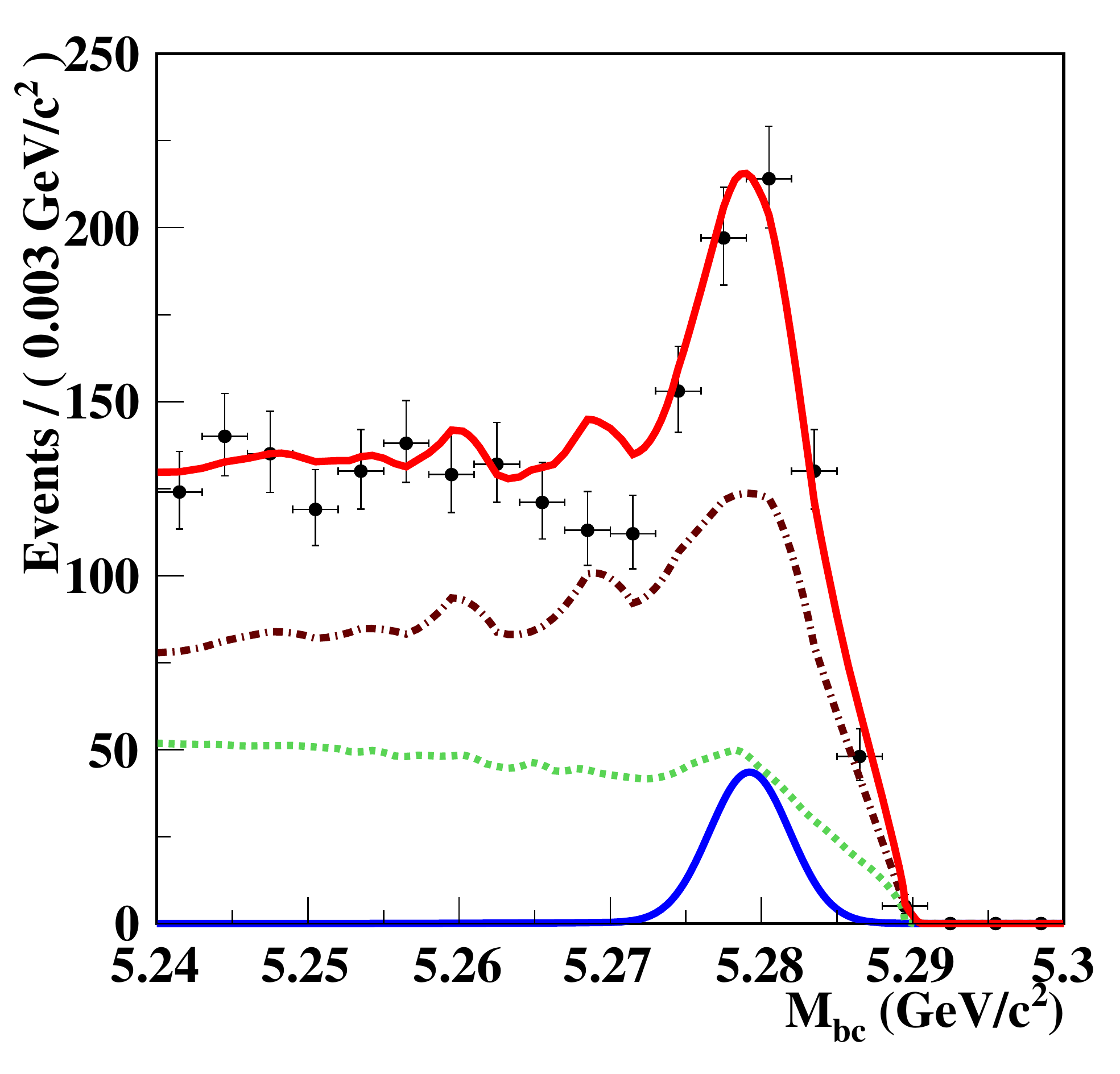}}
\subfigure[$2.5 < \mkk < 3.5$ GeV/$c^2$]{\includegraphics[width=0.24\textwidth]{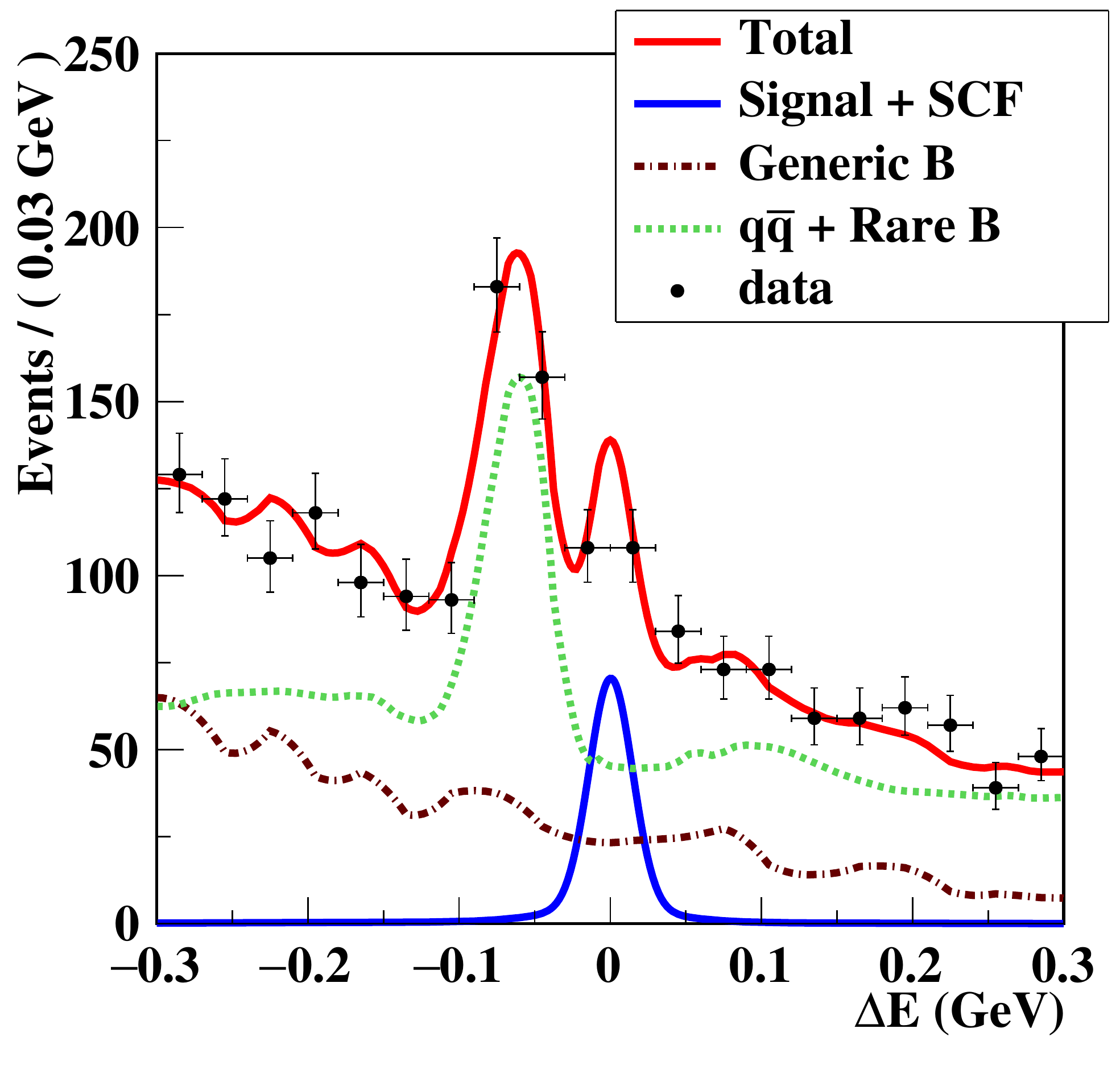}\includegraphics[width=0.24\textwidth]{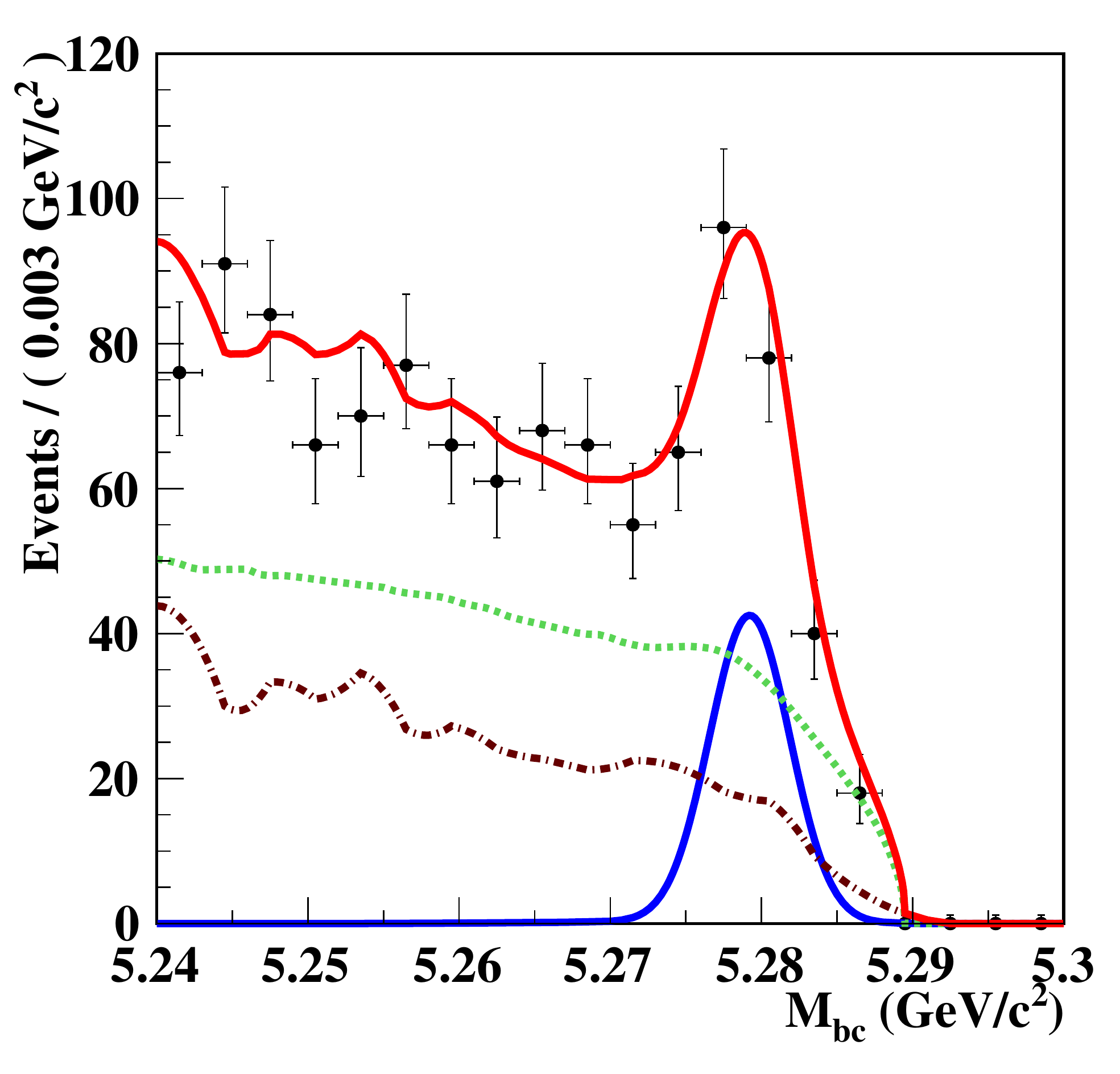}}
\subfigure[$3.5 < \mkk < 5.3$ GeV/$c^2$]{\includegraphics[width=0.24\textwidth]{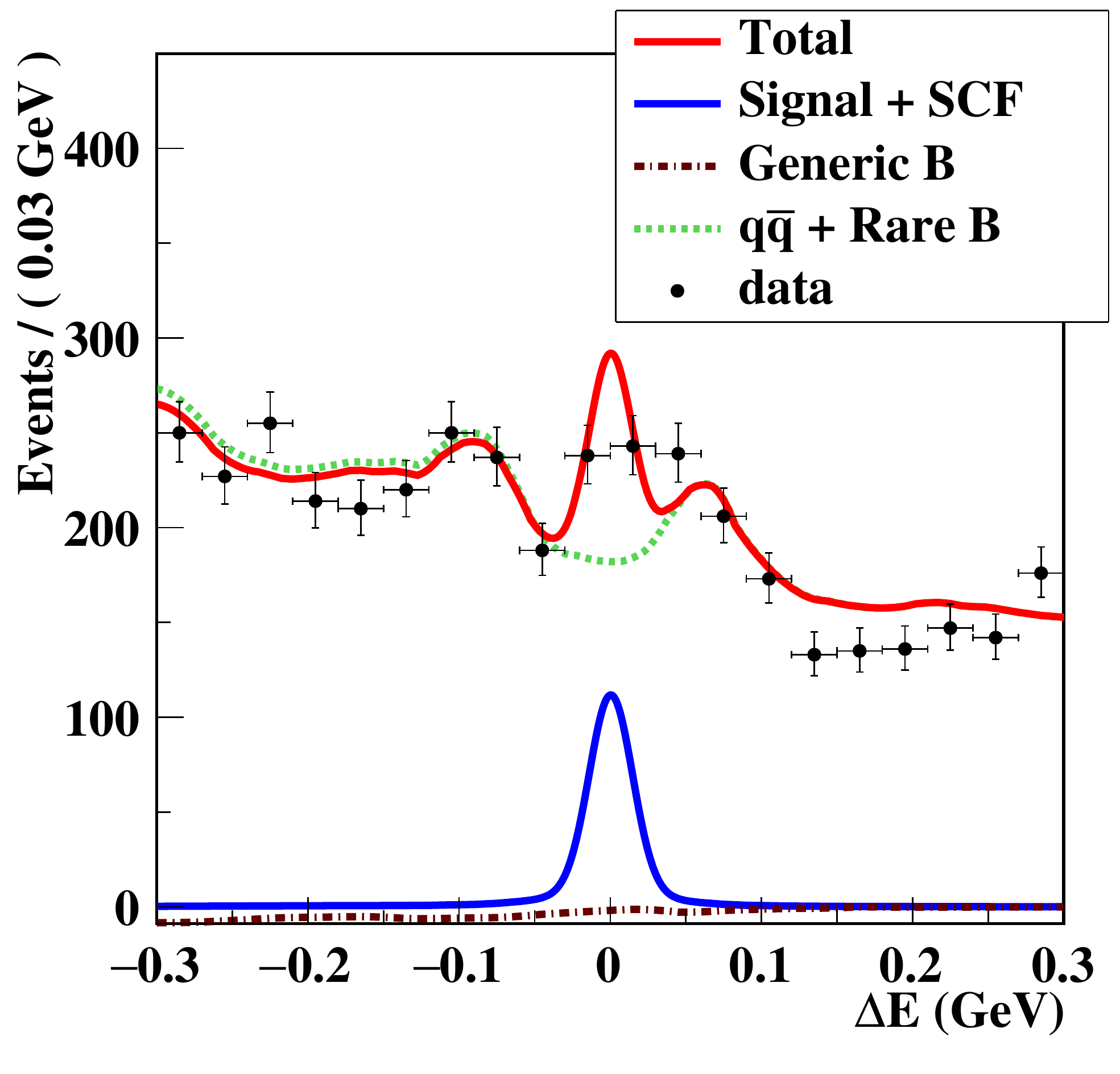}\includegraphics[width=0.24\textwidth]{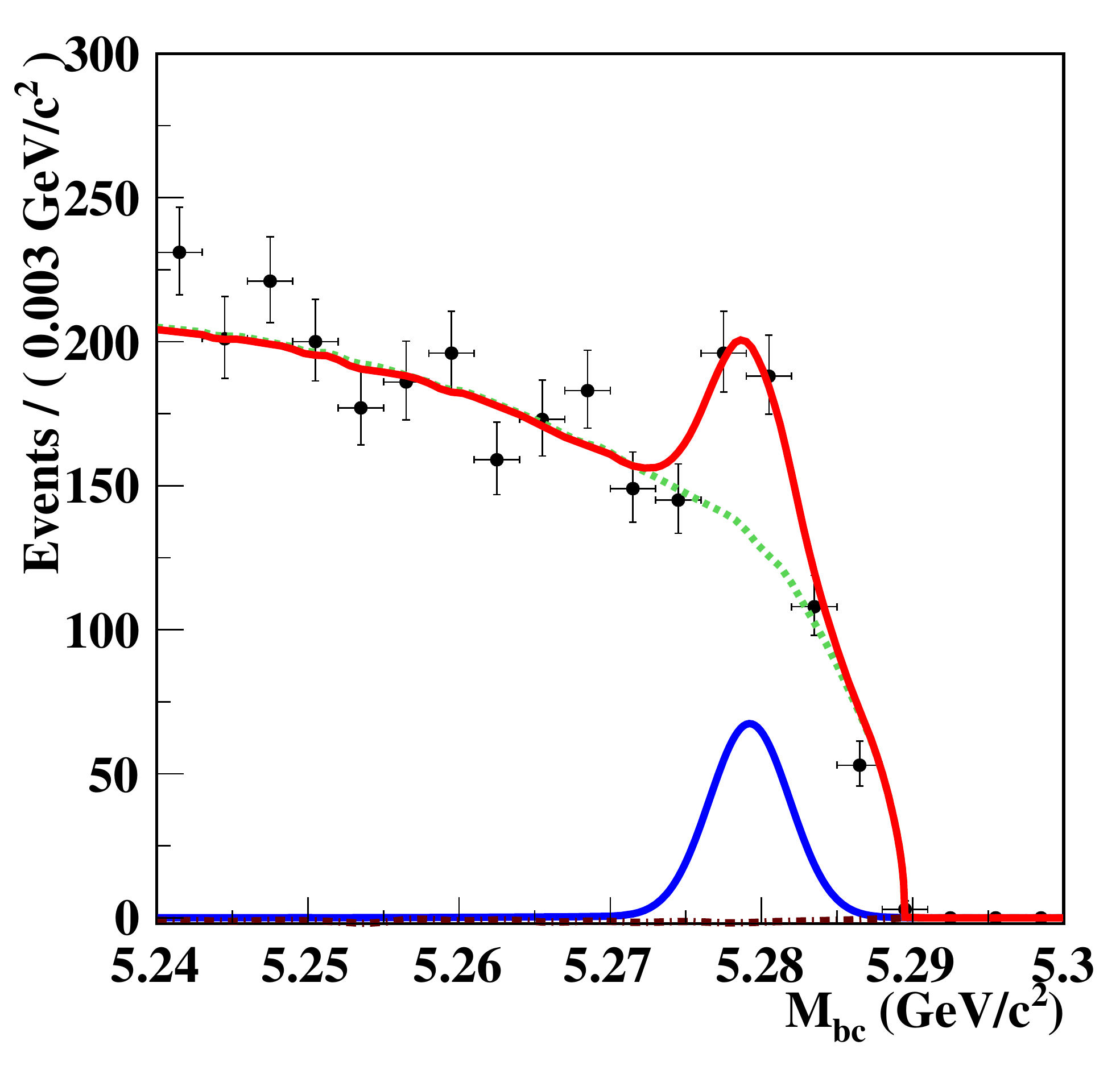}}
\caption{Signal-enhanced projections of the $\mb$-$\de$ fit to data in the $\mkk$ bins. Points with error bars are the data, the red line is the fit result, the blue line is the sum of the signal and the self cross-feed, 
\textcolor{black}{the brown dot-dashed line is the generic $B$ backgrounds, and the green dotted line is the sum of the continuum and the rare $B$ backgrounds.} The projection on $\de$ is with the requirement of $5.275 < \mb < 5.2835$~GeV/$c^2$, while the projection on $\mb$ is with the requirement of $-0.03<\de<0.03$~GeV.
}
\label{fig:app_mkk}
\end{figure}

\clearpage

\section{Fit results in \mkpi bins\label{app:mkpi}}
\begin{figure}[!ht]
\subfigure[$1.0<\mkpi<1.5$ GeV/$c^{2}$]{\includegraphics[width=0.24\textwidth]{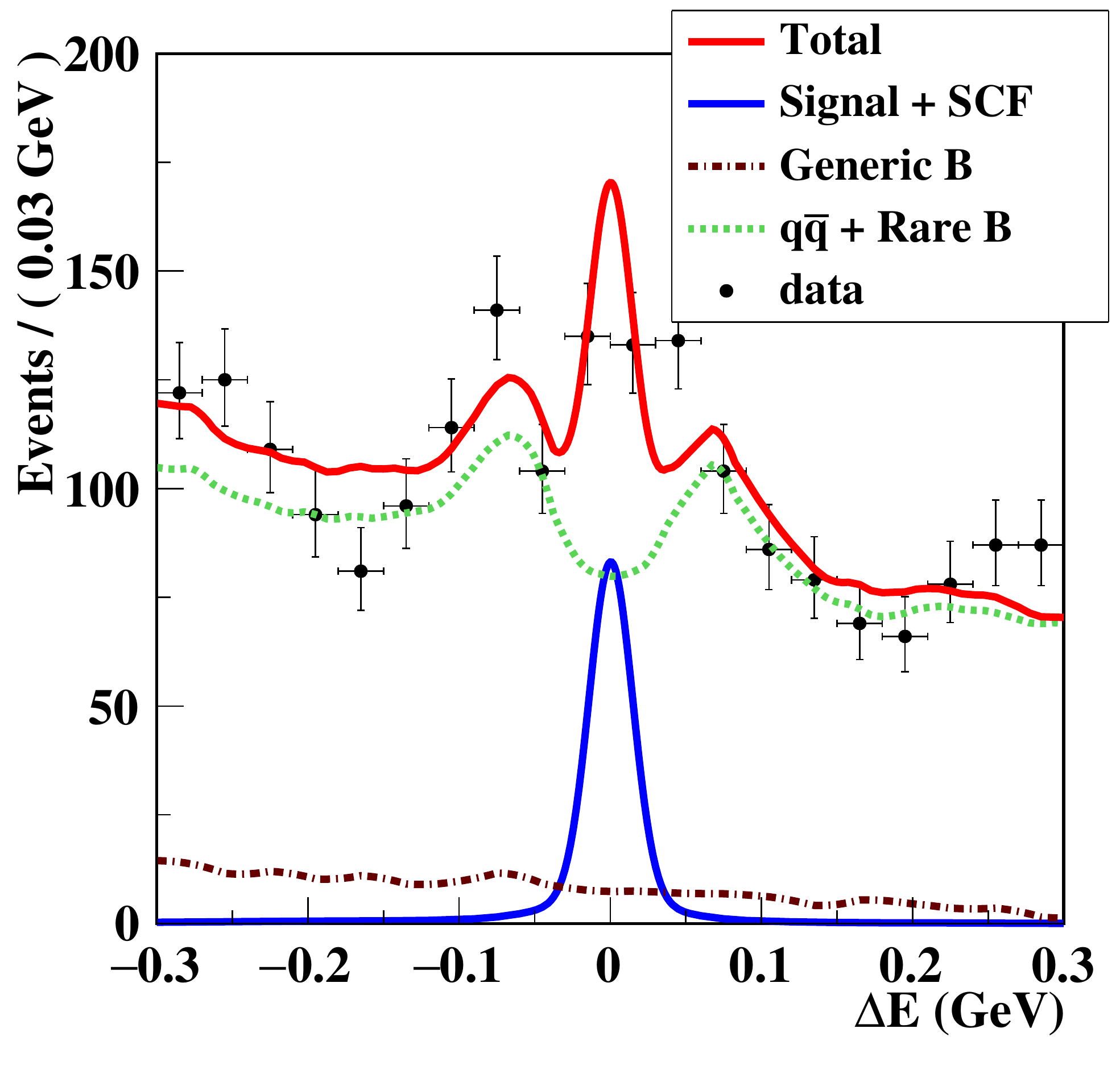}\includegraphics[width=0.24\textwidth]{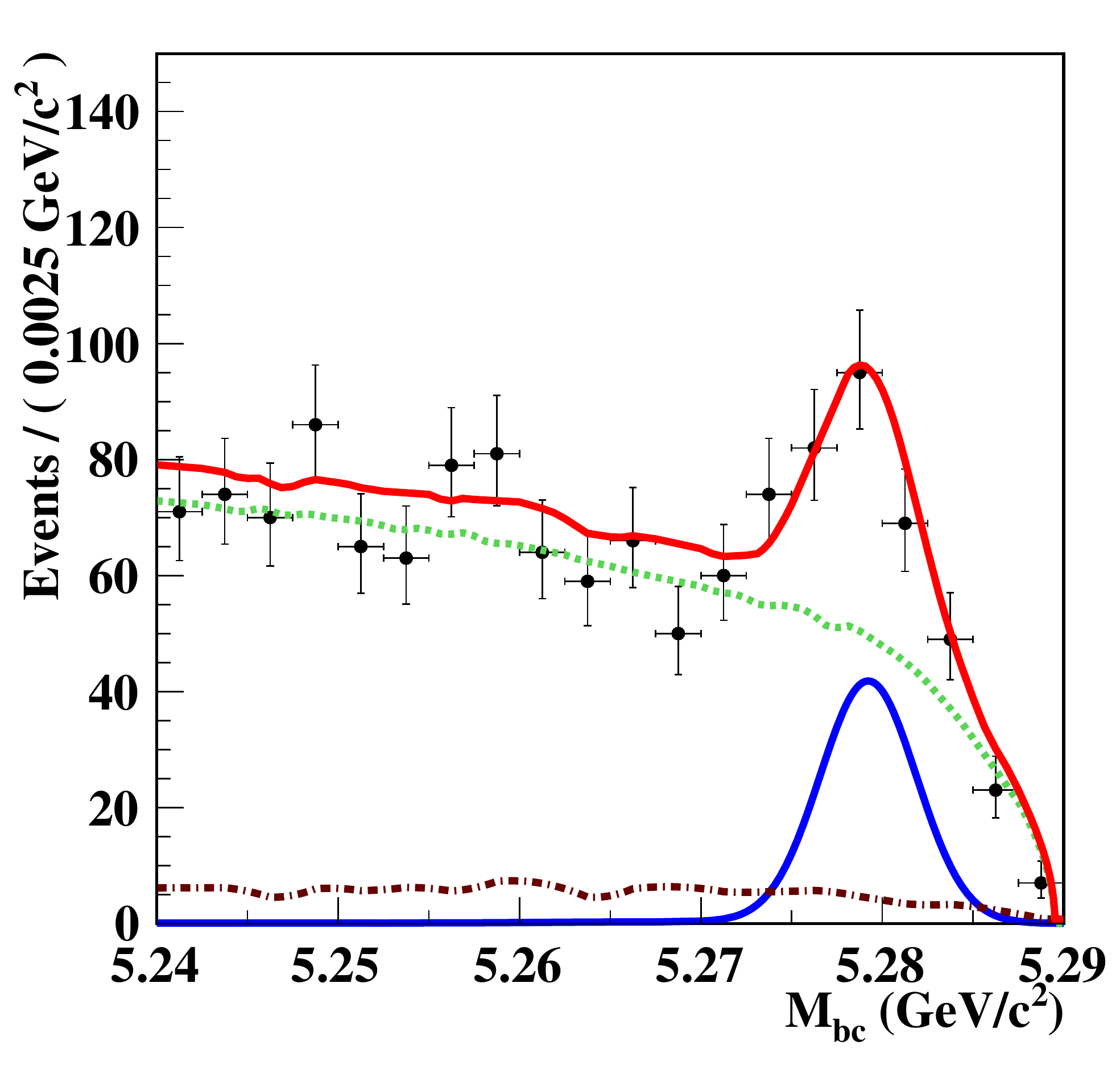}}
\subfigure[$1.5<\mkpi<2.0$ GeV/$c^{2}$]{\includegraphics[width=0.24\textwidth]{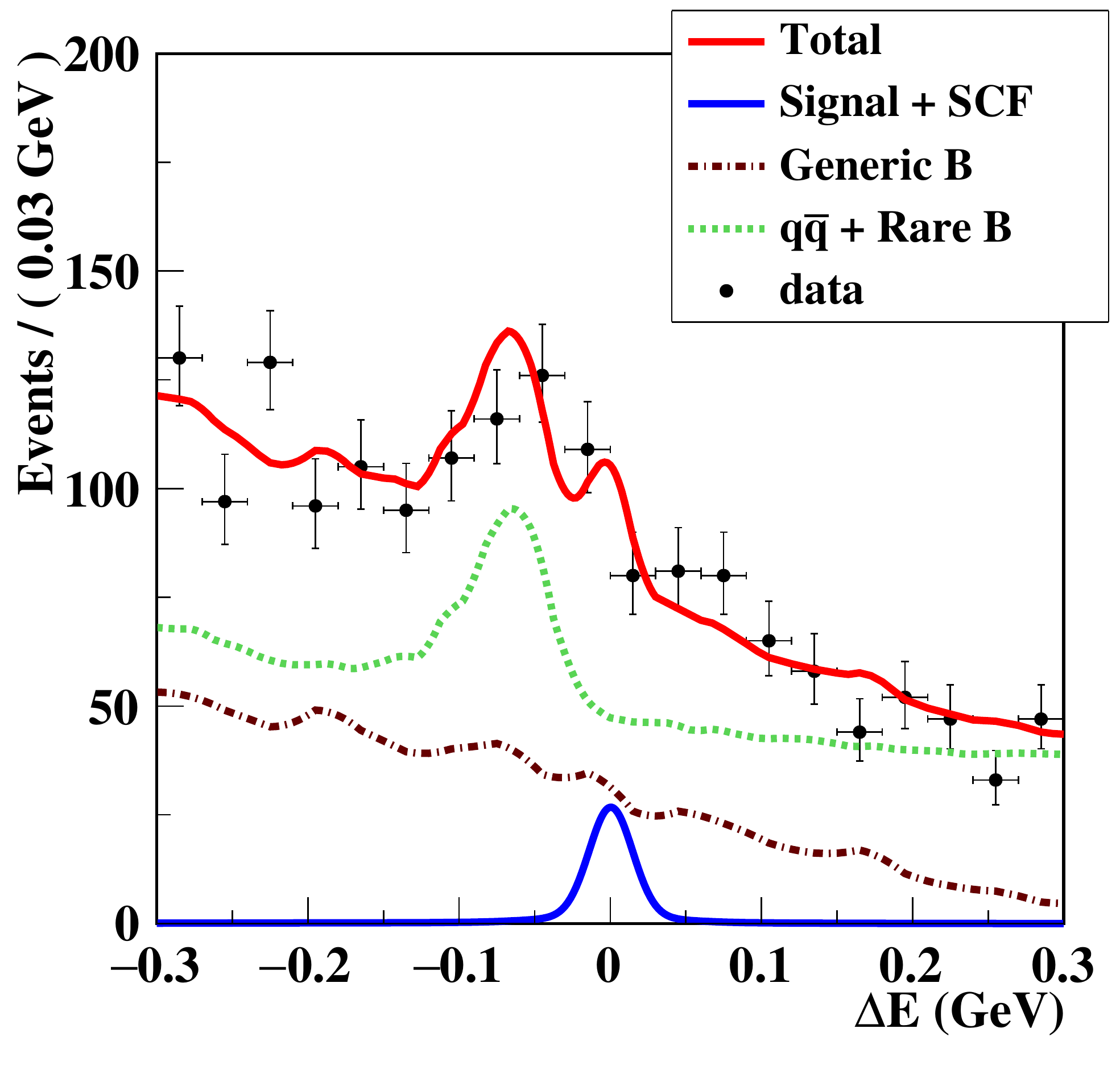}\includegraphics[width=0.24\textwidth]{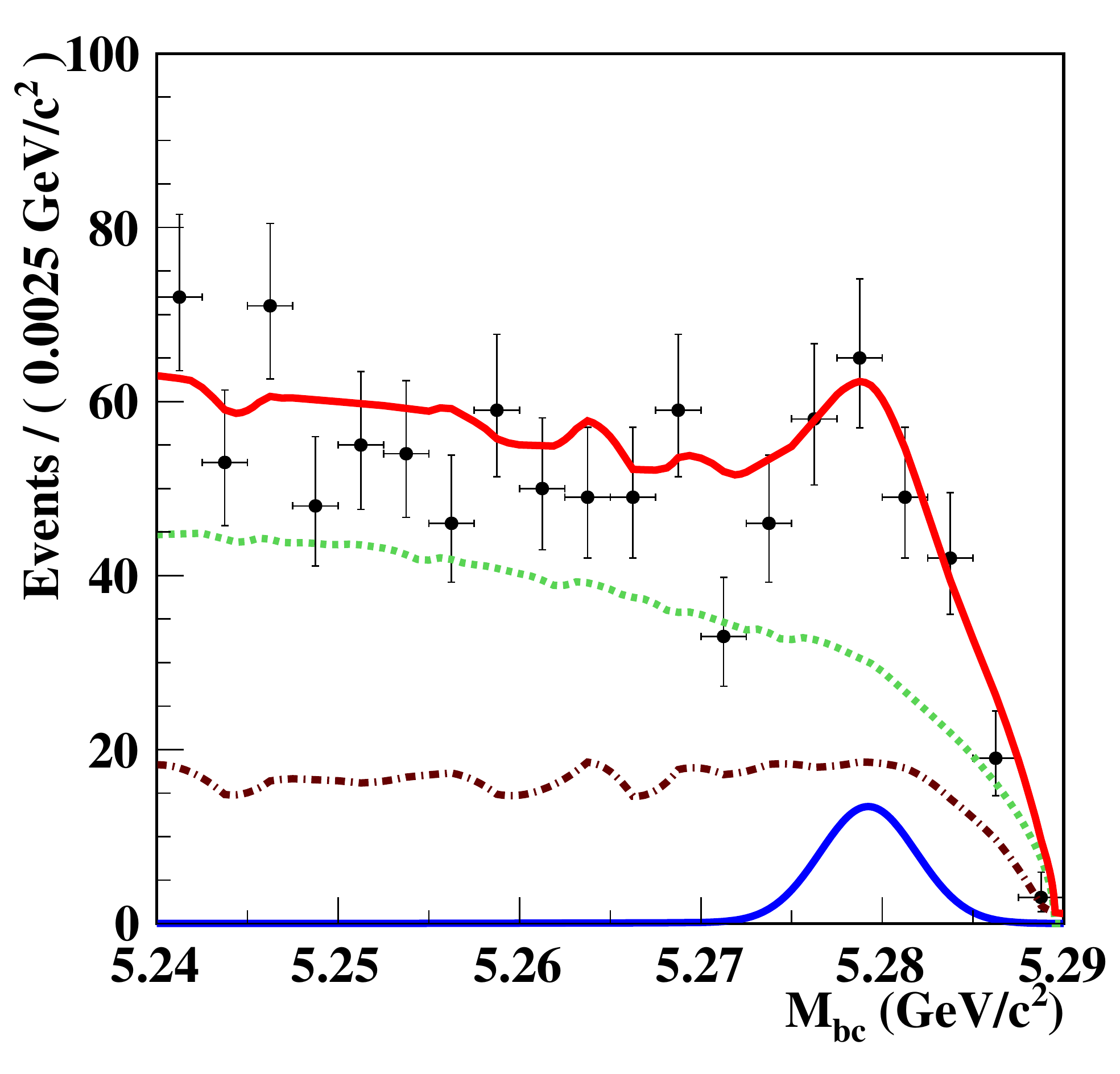}}
\subfigure[$2.0<\mkpi<2.5$ GeV/$c^{2}$]{\includegraphics[width=0.24\textwidth]{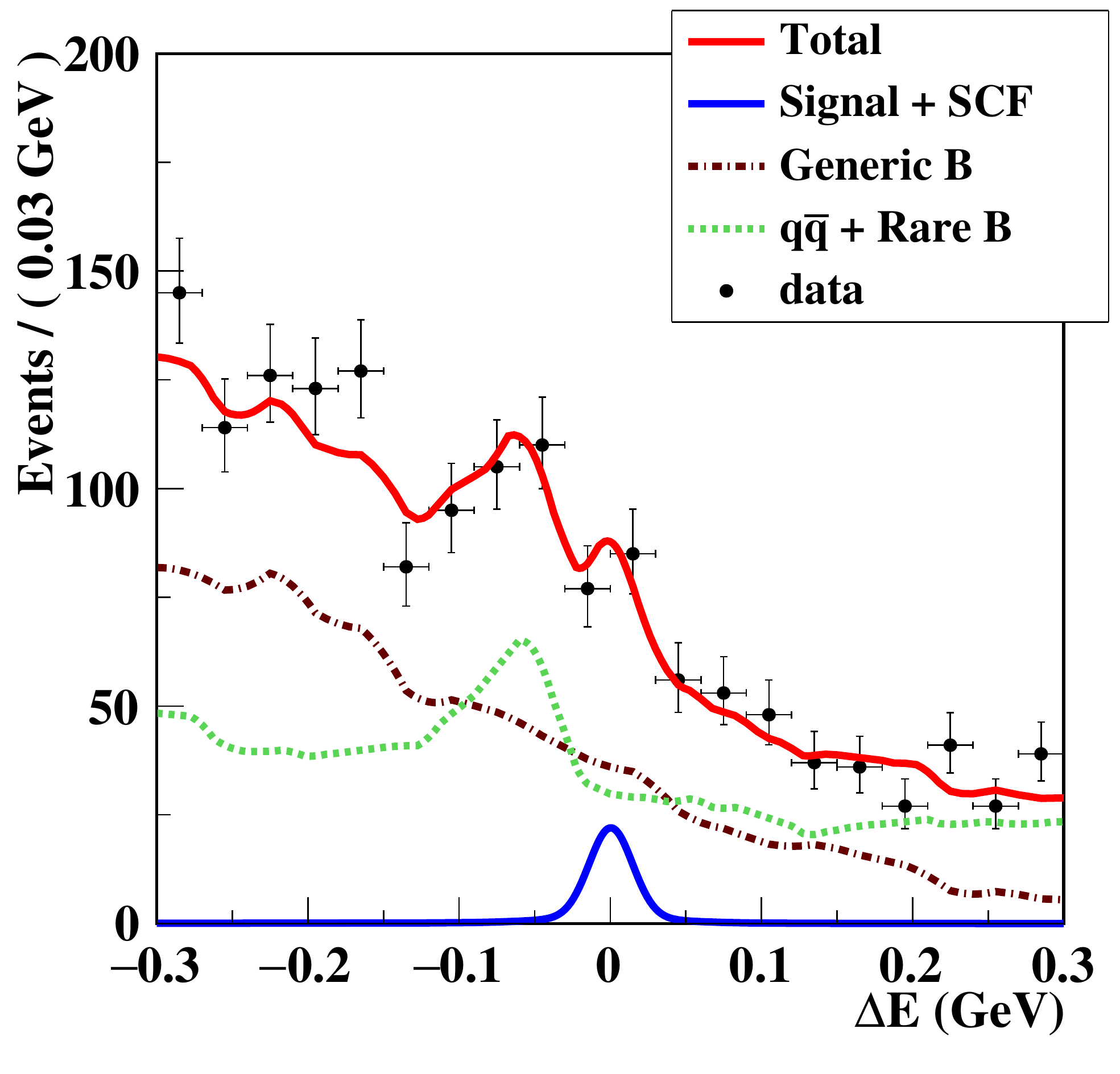}\includegraphics[width=0.24\textwidth]{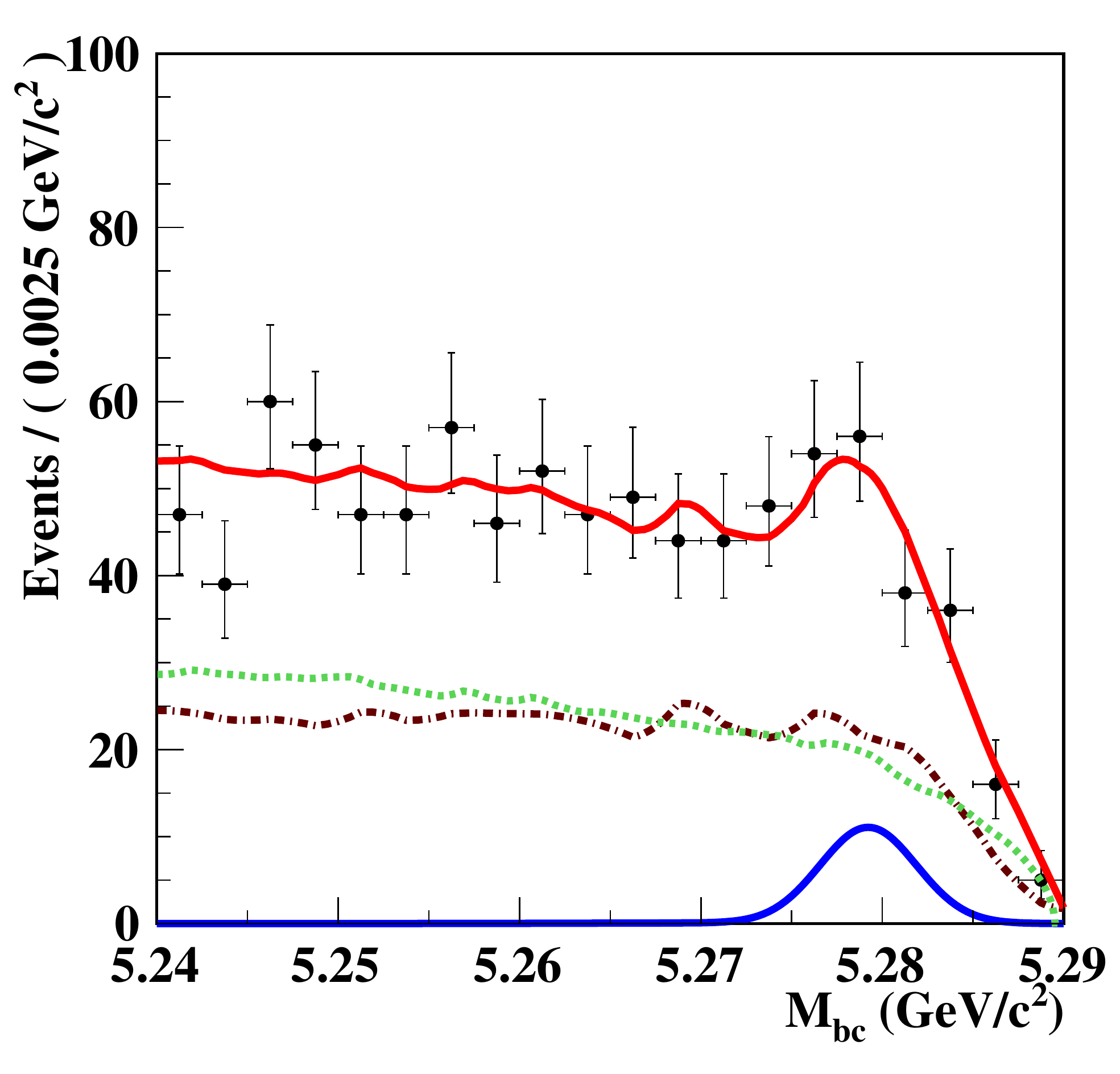}}
\subfigure[$2.5<\mkpi<3.0$ GeV/$c^{2}$]{\includegraphics[width=0.24\textwidth]{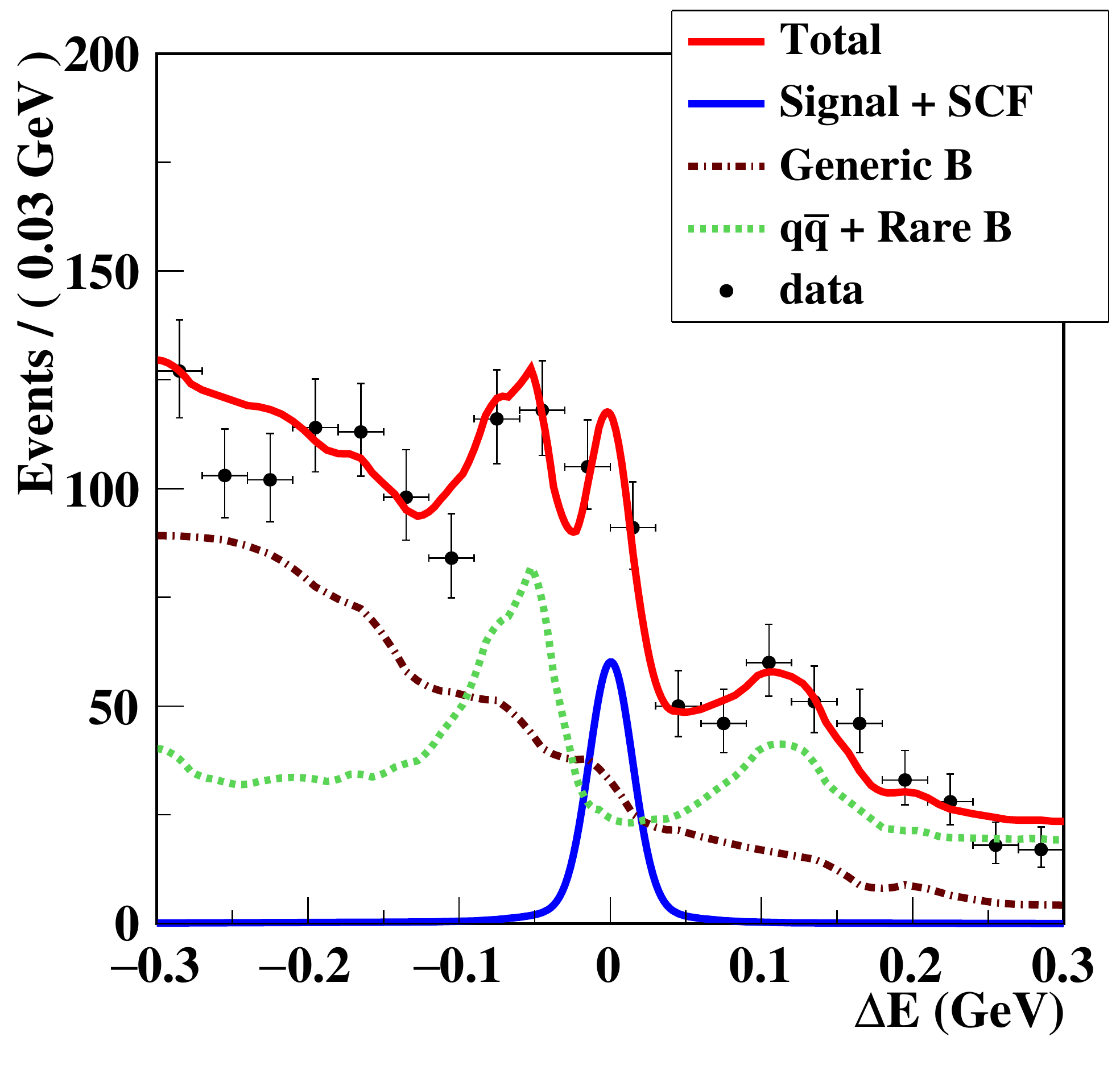}\includegraphics[width=0.24\textwidth]{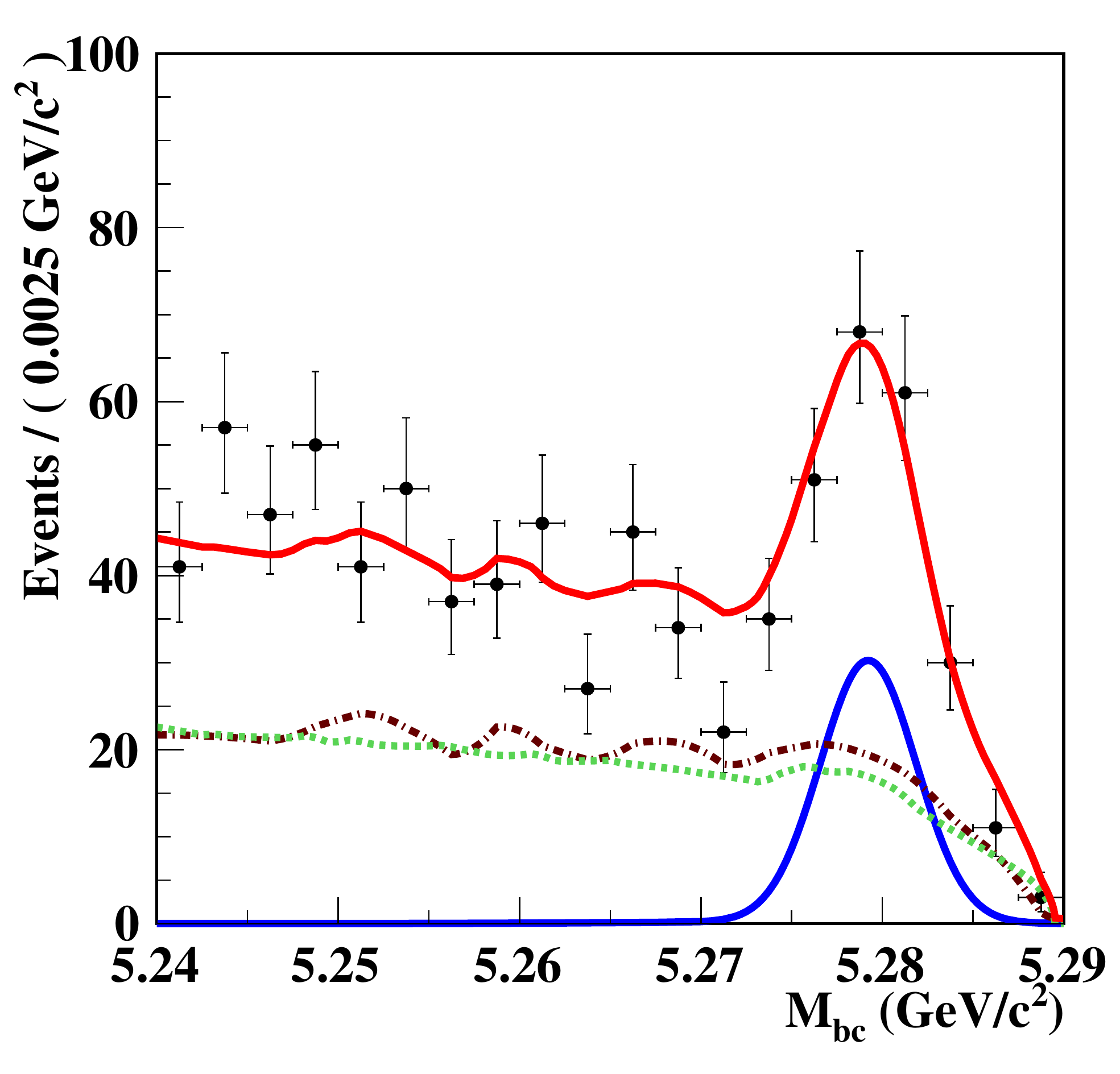}}
\subfigure[$3.0<\mkpi<3.5$ GeV/$c^{2}$]{\includegraphics[width=0.24\textwidth]{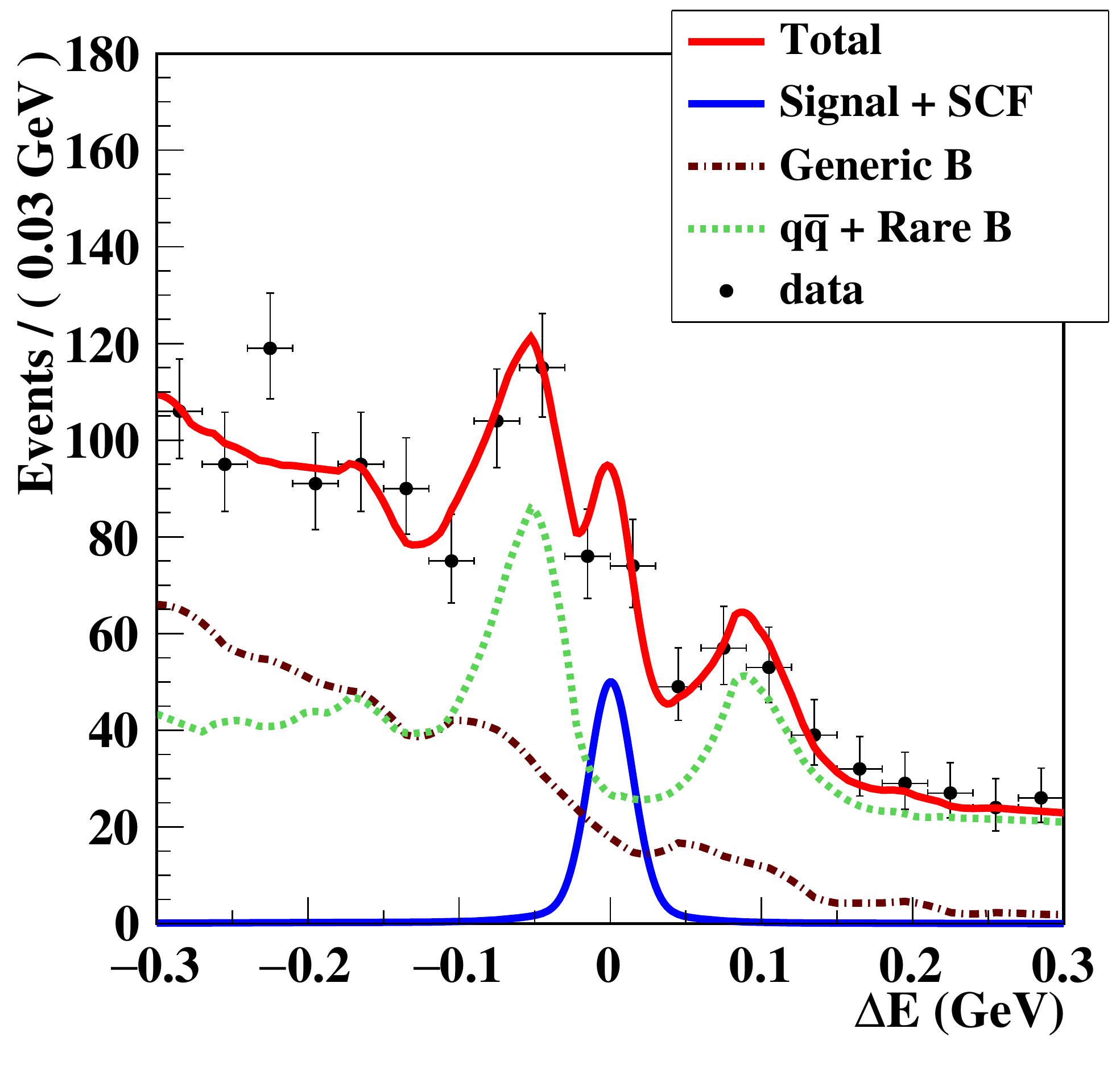}\includegraphics[width=0.24\textwidth]{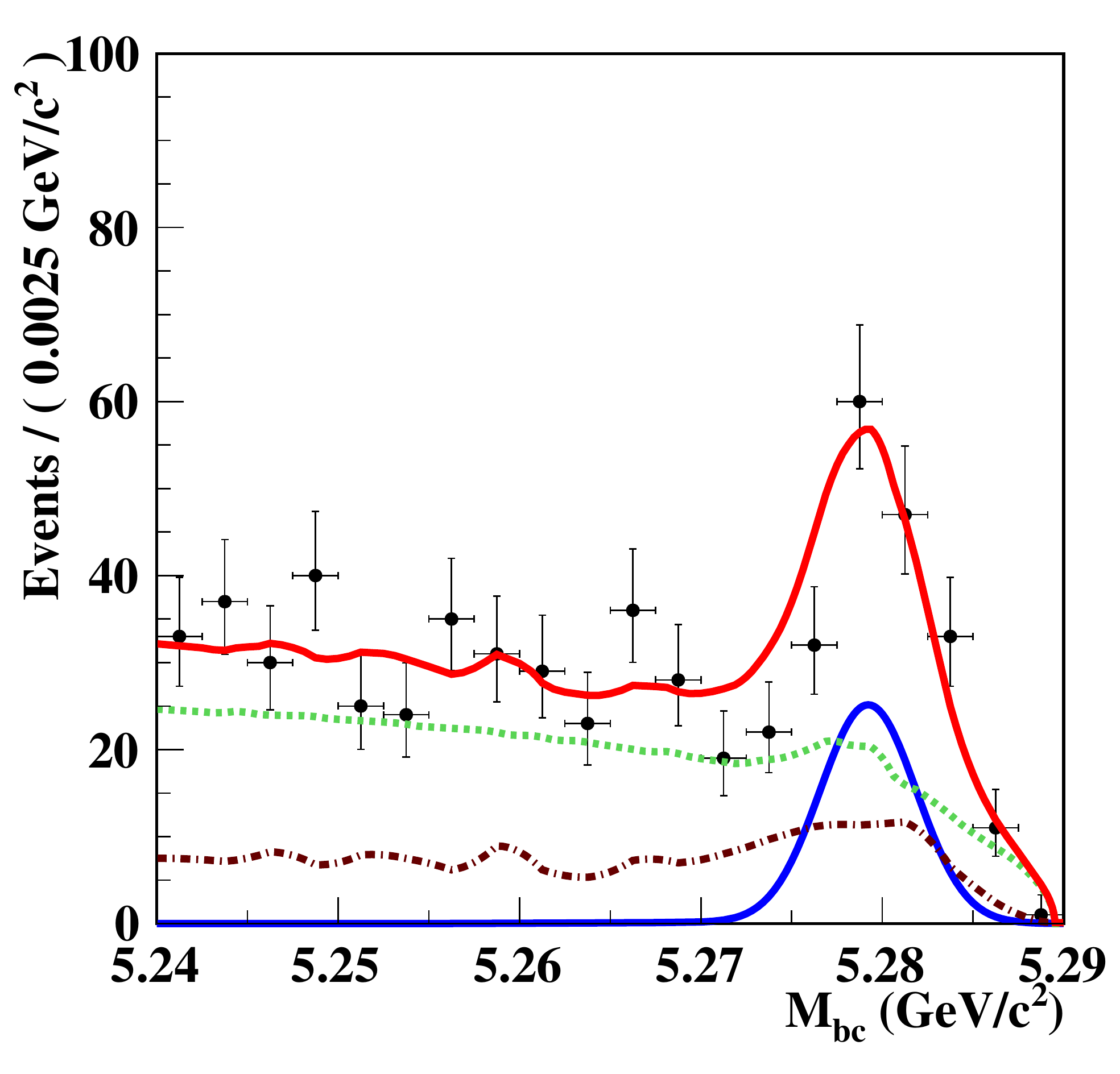}}
\subfigure[$3.5<\mkpi<4.0$ GeV/$c^{2}$]{\includegraphics[width=0.24\textwidth]{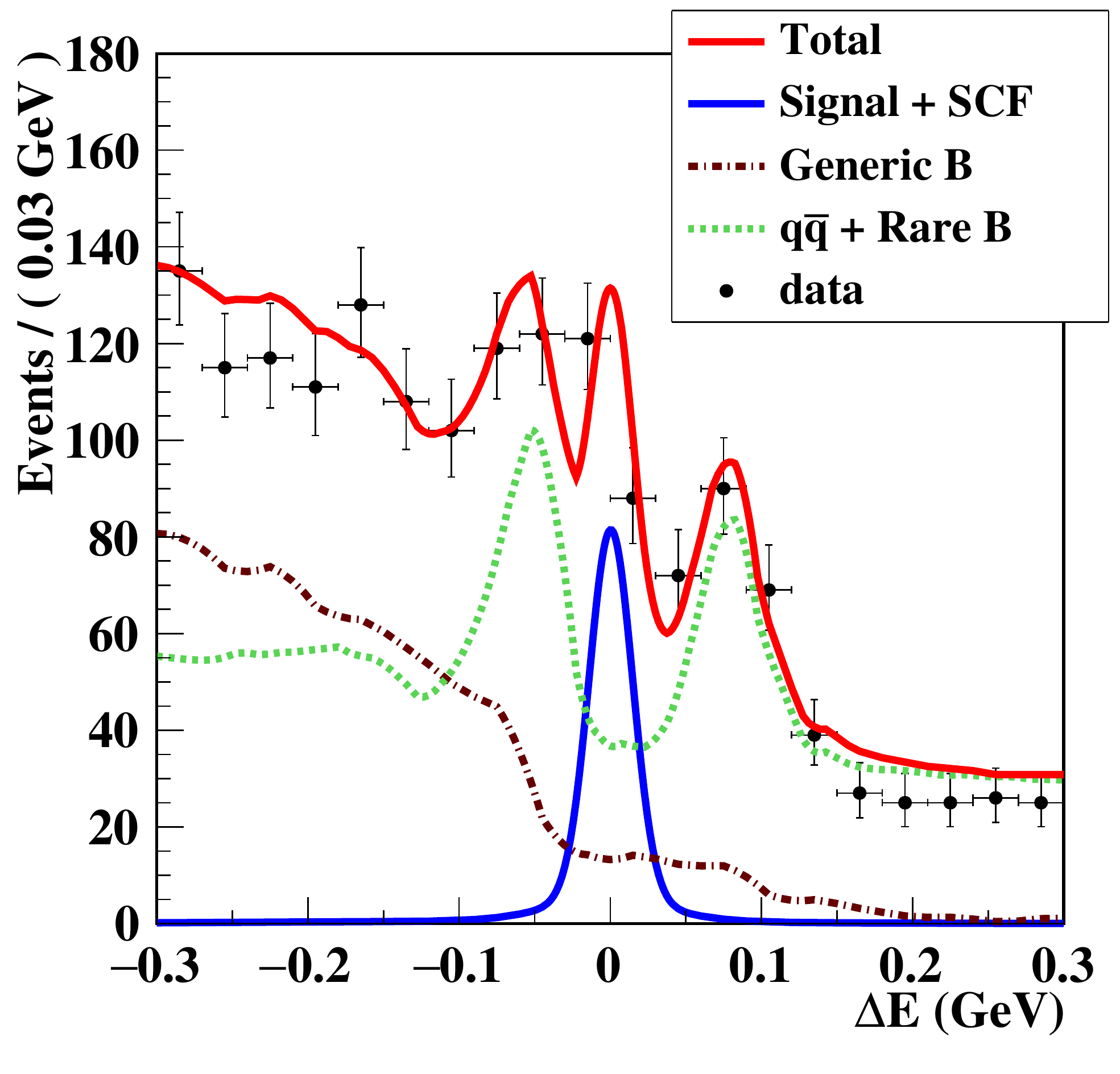}\includegraphics[width=0.24\textwidth]{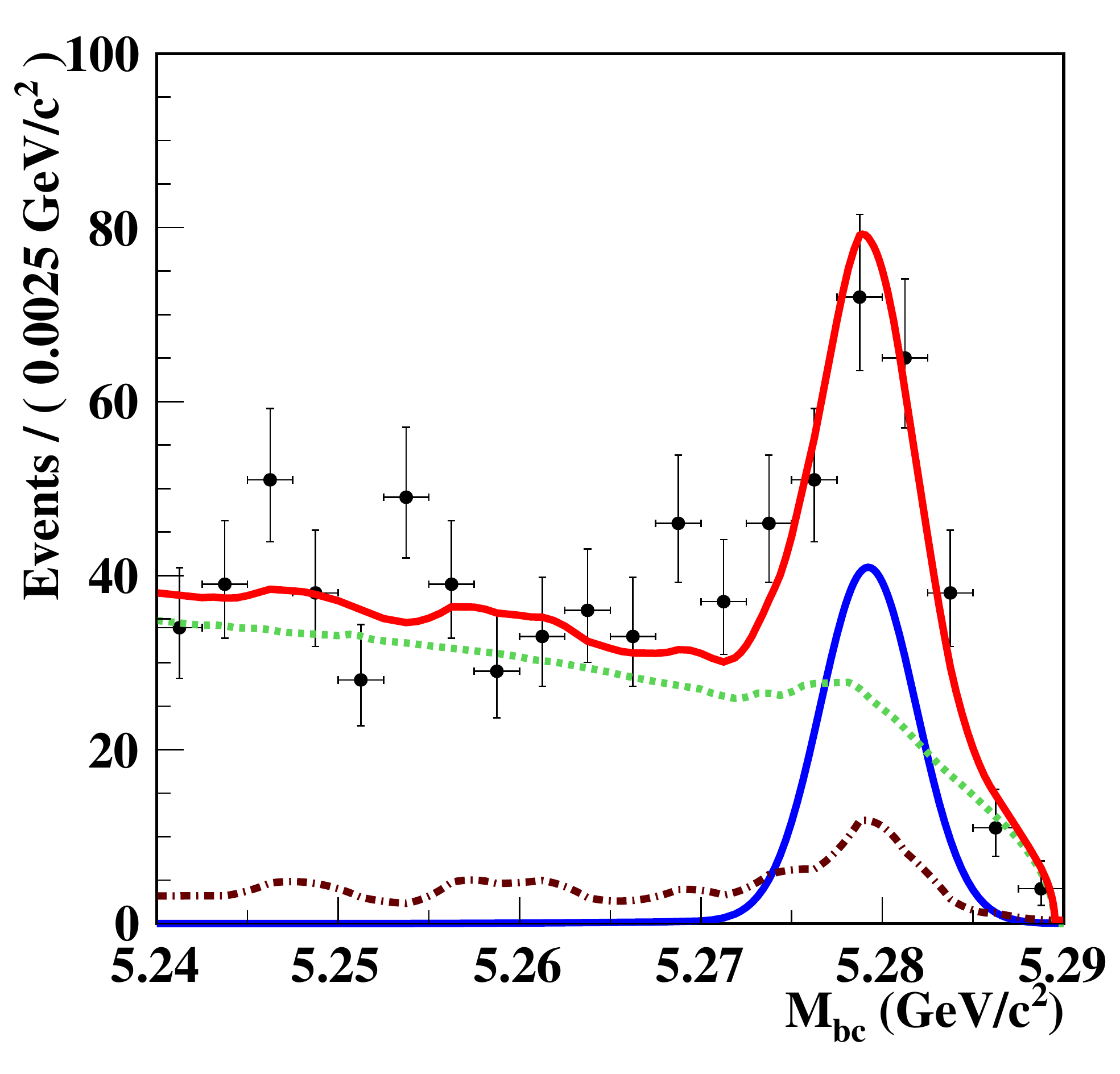}}
\subfigure[$4.0<\mkpi<4.5$ GeV/$c^{2}$]{\includegraphics[width=0.24\textwidth]{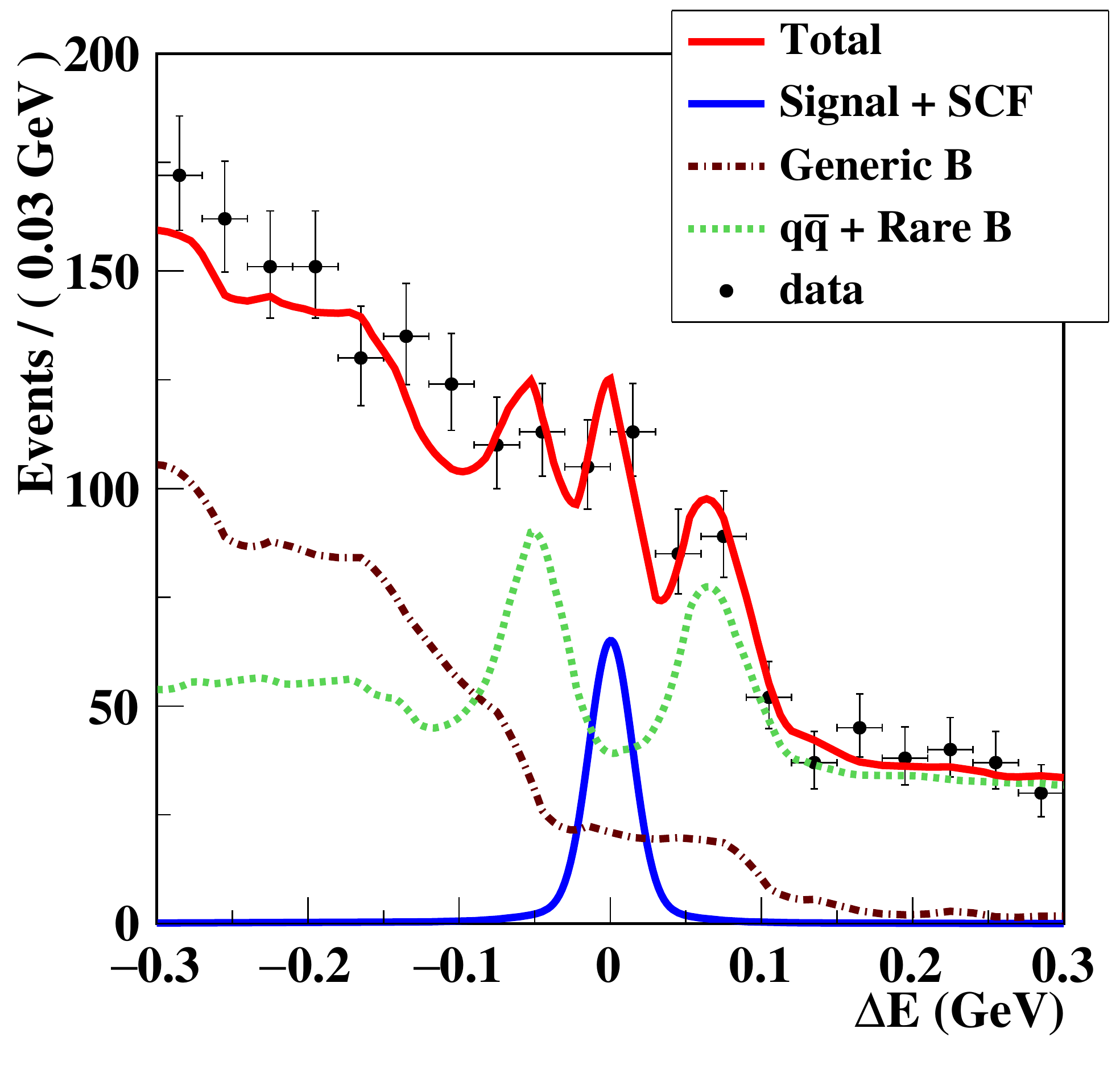}\includegraphics[width=0.24\textwidth]{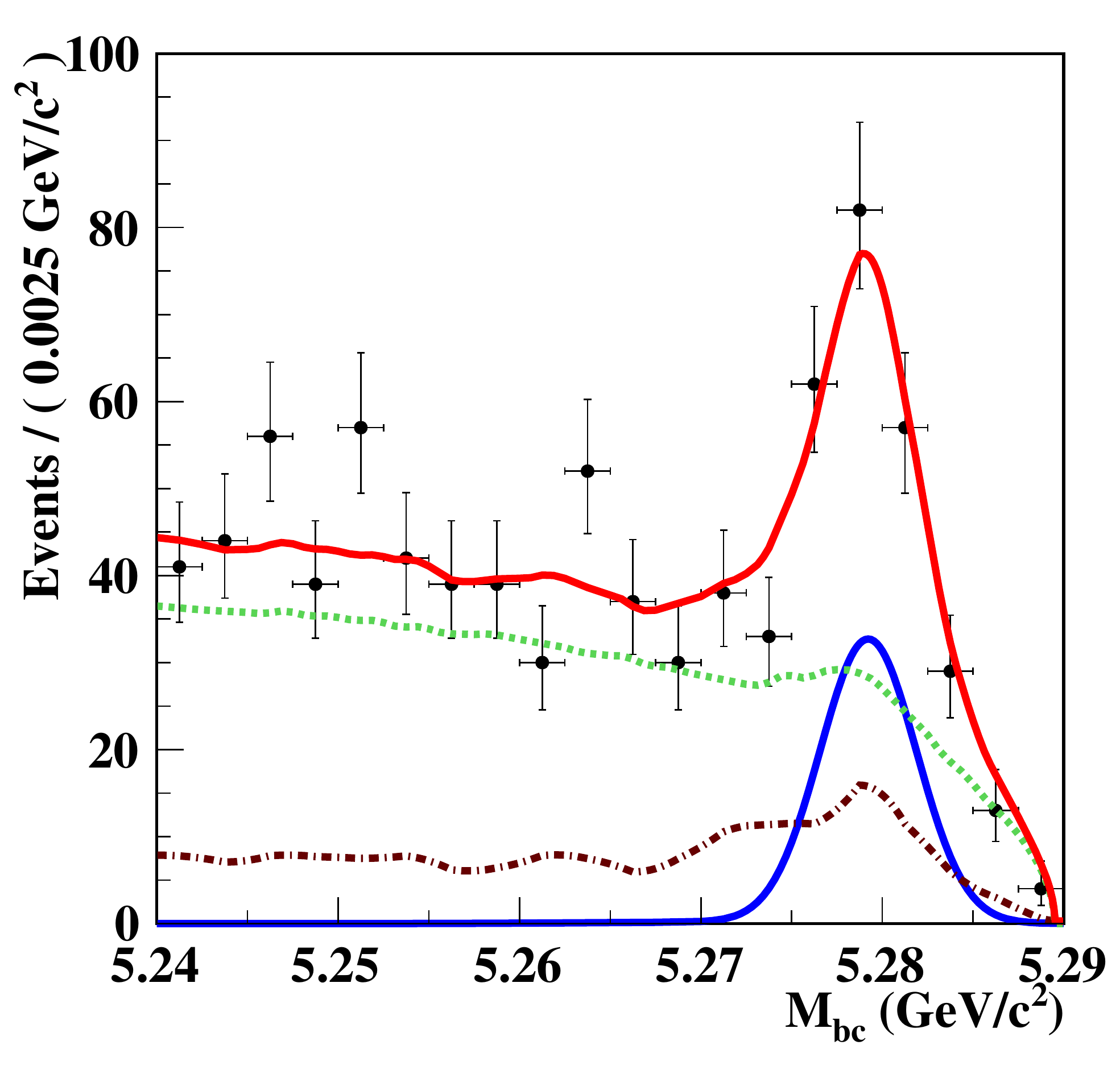}}
\subfigure[$4.5<\mkpi<5.0$ GeV/$c^{2}$]{\includegraphics[width=0.24\textwidth]{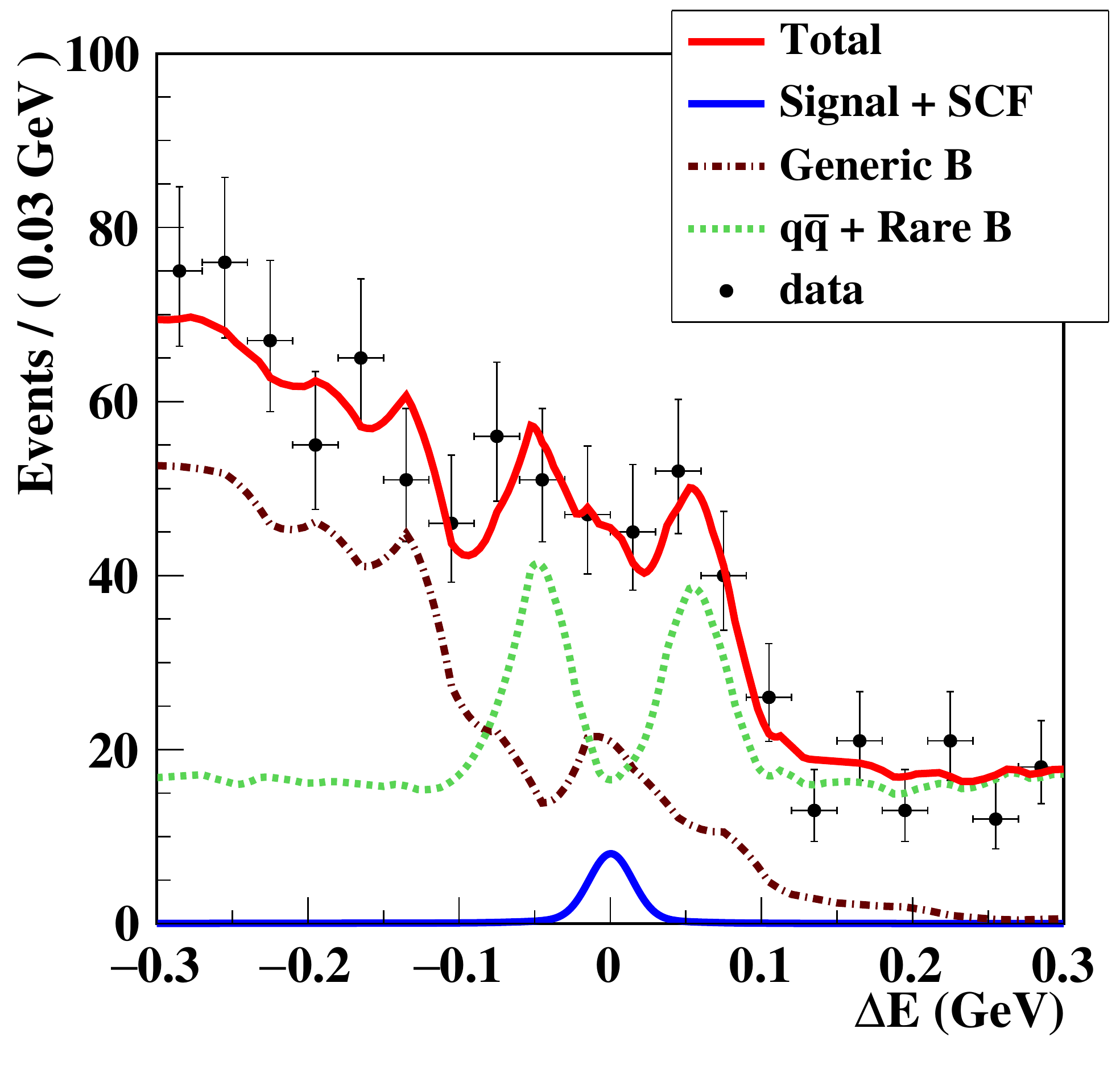}\includegraphics[width=0.24\textwidth]{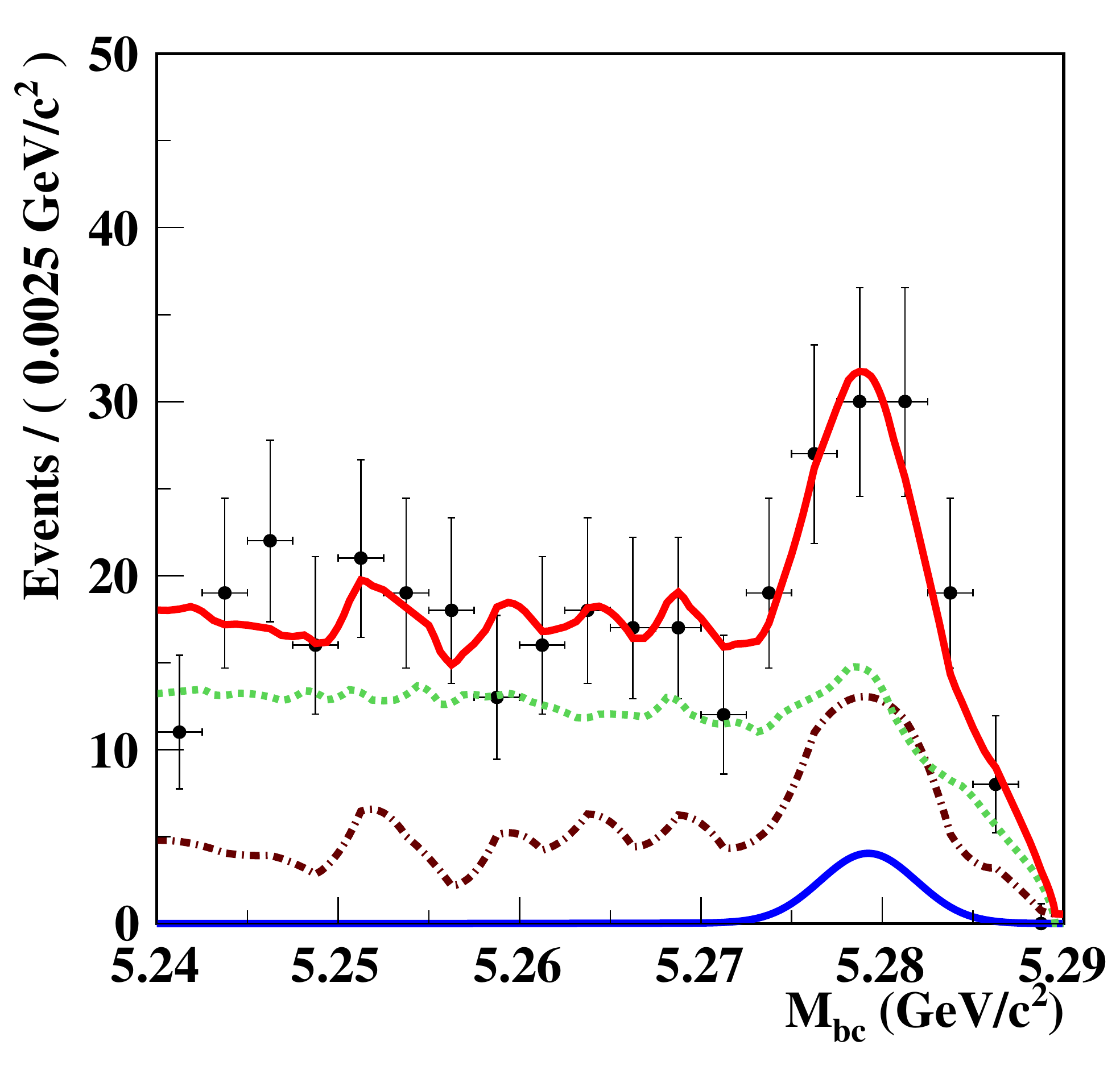}}
\caption{Signal-enhanced projections of the $\mb$-$\de$ fit to data in the $\mkpi$ bins. Points with error bars are the data, the red line is the fit result, the blue line is the sum of the signal and the self cross-feed, \textcolor{black}{the brown dot-dashed line is the generic $B$ backgrounds, and the green dash-dot line is the sum of the continuum and the rare $B$ backgrounds.} The projection on $\de$ is with the requirement of $5.275 < \mb < 5.2835$~GeV/$c^2$, while the projection on $\mb$ is with the requirement of $-0.03<\de<0.03$~GeV.}
\label{fig:app_mkpi}
\end{figure}

\twocolumngrid


\begin{thebibliography}{99}
\bibitem{conjugate} The inclusion of charge-conjugate modes is implied throughout this paper.
\end{thebibliography}
\end{document}